\documentclass[prc,showpacs,twocolumn]{revtex4}
\usepackage{graphicx}
\usepackage{dcolumn}
\usepackage{bm}

\begin{document}

\title{Microscopic description of fission in  Uranium isotopes 
with the Gogny energy density functional}

\author{R. Rodr\'{\i}guez-Guzm\'an}

\email{raynerrobertorodriguez@gmail.com}

\affiliation{Department of  Physics and Astronomy, Rice University, 
Houston, Texas 77005, USA}

\affiliation{Department of Chemistry, Rice University, Houston, Texas 77005, USA}

\author{L.M. Robledo}

\affiliation{Departamento  de F\'{\i}sica Te\'orica, 
Universidad Aut\'onoma de Madrid, 28049-Madrid, Spain}

\email{luis.robledo@uam.es}

\date{\today}

\begin{abstract}
The most recent parametrizations D1S, D1N and D1M  of the Gogny energy density 
functional are used to describe fission in the isotopes $^{232-280}$
U. Fission paths, collective masses and zero point  quantum 
corrections, obtained within the constrained Hartree-Fock-Bogoliubov 
approximation, are used to compute the systematics of the 
spontaneous fission half-lives $t_\mathrm{SF}$, the masses and 
charges of the fission fragments as well as their intrinsic shapes.
The Gogny-D1M parametrization has been benchmarked against available
experimental data on inner and  second barrier heights, excitation 
energies of the fission isomers and half-lives in a selected set of 
 Pu, Cm, Cf, Fm, No, Rf, Sg, Hs and Fl nuclei. It is concluded that
 D1M represents a reasonable starting point to describe fission in heavy 
and superheavy nuclei. 
Special attention is also paid to understand the uncertainties in the 
predicted $t_\mathrm{SF}$ values  arising from the different 
building blocks entering the standard semi-classical 
Wentzel-Kramers-Brillouin formula. Although the uncertainties
are large, the trend with mass or neutron numbers are well reproduced
and therefore the theory still has predictive power. In this respect, it 
is also shown that modifications of a few per cent in the pairing strength
can have a significant impact on the collective masses leading to uncertainties in 
the $t_\mathrm{SF}$ values of several orders of magnitude.

\end{abstract}


\maketitle{}

% ----------------------------------------------------------------------
%
%  S E C T I O N
%
%                                      I n t r o d u c t i o n  
%
% ----------------------------------------------------------------------  

\section{Introduction.}

Nuclear fission is, at the same time, one of the most distinctive phenomenon in the
physics of the nucleus and one of the most elusive to a theoretical description.
It takes place mostly in heavy and superheavy
nuclei and involves the evolution of the initial parent system from its ground state 
to scission through a sequence of intrinsic 
shapes labeled by some sort of deformation parameter \cite{Specht,Bjor,Krappe}. Once 
the scission configuration is reached, the system splits in two daughter nuclei. The 
occurrence of fission is the result of the competition between the nuclear
surface energy  coming from the strong  interaction and the Coulomb 
repulsion of the nuclear charge density \cite{MeitnerFrisch}.  
In fact, nuclear fission was originally described \cite{MeitnerFrisch,Bohr-Wheeler-fission}
in terms of the liquid-drop model where the surface tension plays an essential role. 
However, experimental and 
theoretical evidences  emphasize the  stabilizing role 
of shells effects \cite{rs,Lawrence-Wheeler,Strutisnky-1} and therefore much 
effort has been laid on the development of models that
incorporate those effects to the semi-classical liquid-drop model
description \cite{Strutisnky-ShellCorr-Method-1,Strutisnky-ShellCorr-Method-2,Myers-Swiatecki}. 
The outcome of these models (see, for example, Refs. \cite{Moller-1,Moller-2}
and references therein) is a potential energy surface, expressed in terms of 
several deformation parameters, showing a quite involved topography (direct
consequence of shell effects) with minima, valleys, ridges and saddle points.
In this picture, fission is the journey along this complicated landscape from the
ground state to the scission point (an elusive concept to be discussed later). 
In spite of their success in describing some fission observables, these models
lack essential quantum mechanisms like  tunneling through a classically forbidden
barrier or a sound description of the inertia associated to the collective degrees 
of freedom used to describe fission.

From a more fundamental point of view, fission could be regarded as a quantum
mechanical problem describing the evolution from some given initial quantum state to
a final state with two fragments and involving tunneling through a potential
barrier defined in a multidimensional space. The initial state can be 
the parent nucleus ground state in spontaneous fission or a highly excited state
(usually described as a statistical admixture by assuming thermal equilibrium) 
in induced fission. Although several attempts to deal with
this problem in a path-integral framework involving instantons and other
sophisticated concepts have been considered \cite{negele,skalski} in the
past, it has not been possible to establish a computationally feasible framework
capable to describe real nuclei with realistic interactions. Therefore, it is
customary to use a 
more phenomenological approach where the dynamical changes involved in the transition 
from a single nucleus to two fragments are usually described in the framework
of the (constrained) self-consistent mean-field approximation \cite{rs,Bender-review}
based on a given non-relativistic Energy Density Functional (EDF) 
of the Gogny  
\cite{gogny-d1s,Delaroche-2006,Robledo-Martin,Dubray,PEREZ-ROBLEDO,Younes2009,Warda-Egido-Robledo-Pomorski-2002,Egido-other1,
Warda-Egido-2012}
and/or Skyrme 
\cite{UNEDF1,Mcdonell-1,Mcdonell-2,Erler2012,Baran-SF-2012,Baran-1981}
type as well as with several parametrizations of the relativistic mean-field (RMF) Lagrangian 
\cite{Abusara-2010,Abu-2012-bheights,RMF-LU-2012,Kara-RMF,Afa-arxiv}. Recently, the 
fission properties of the Barcelona-Catania-Paris-Madrid (BCPM) EDF \cite{BCPM} have also been studied
in the case of dripline-to-dripline Uranium isotopes \cite{Robledo-Giulliani}. 
The aim of all these methods is to determine the relevant fission configurations by
means of  the Hartree-Fock-Bogoliubov (HFB) mean-field method using constraints
on relevant quantities associated with shape parameters like multipole moments
or neck degrees of freedom. 
The resulting HFB wave functions are then used to compute other parameters
like the collective inertia and quantum corrections to the potential energy
surface stemming from the restoration of broken symmetries (rotational, parity, etc)
and fluctuations in the collective parameters defining the fission paths.
An implicit assumption of this framework is that the fission properties are 
determined by general features of the interactions and therefore no fine-tuning
should be required to describe fission observables. However, interactions
are usually tuned to reproduce nuclear matter parameters that are not
properly constrained by experimental data (the typical case pertaining fission 
is the surface energy of semi-infinite nuclear matter) and therefore there 
are examples of interactions fitted to fission properties like Gogny-D1S 
\cite{gogny-d1s,gogny}, the old SkM* \cite{skm} or the more recent UNEDF1 
\cite{UNEDF1,Mcdonell-1} Skyrme parametrization.
   
Typical fission observables are the fission lifetimes, fragment mass 
distributions and kinetic energy of the fragments. Also fission barrier
heights are commonly considered as experimental "pseudo-data". All those quantities
are required in many physical scenarios like the stability of superheavy 
elements or the final stages of the r-process in stellar nucleosynthesis that
are responsible for most of the abundance of heavy elements in the solar system.     
Fission remains a topic of high current interest not only in several areas 
of basic science but also in the  application's side where the issues of
efficient energy production with nuclear reactors or the degradation of long-lived 
radioactive waste are of great interest \cite{Krappe,Wagemans}. 

It turns out that fission observables are quite sensitive to pairing 
correlations (see \cite{Robledo-Giulliani} for a recent discussion) due
to the strong dependence of the collective inertias with the inverse of
the square of the pairing gap \cite{proportional-1,proportional-2}. They are also
very sensitive to the underlying theory used to describe collective motion
(typically the Adiabatic Time Dependent HFB (ATDHFB) or
the Generator Coordinate Method (GCM)) and the approximations involved
in the evaluation of the inertias (see
\cite{crankingAPPROX,Giannoni,Libert-1999} for different approximations).
As a consequence, fission can be considered as a very demanding testing-ground
for theories and interactions used in nuclear structure calculations.

In the  last decades there has been a renewed interest in microscopic fission studies 
\cite{Delaroche-2006,Robledo-Martin,Dubray,PEREZ-ROBLEDO,Younes2009,Warda-Egido-Robledo-Pomorski-2002,Egido-other1,
Warda-Egido-2012,UNEDF1,Mcdonell-1,Mcdonell-2,Erler2012,Baran-SF-2012,Baran-1981,Abusara-2010,Abu-2012-bheights,RMF-LU-2012,
Kara-RMF,Afa-arxiv,Robledo-Giulliani}
due to the wealth of information in actinide nuclei \cite{Specht}, the huge 
progress in the production of  superheavy elements, via 
cold and hot fusion reactions, and the new possibilities
opened up by heavy-ion collisions with radioactive ion beams
(see, for example, 
Refs. \cite{Hoffman-SHE,Gates-SHE,Oganessian-1,Oganessian-2,Oganessian-3,Haba-SHE,Gerl-RIB,Stavsetra-SHE,JULIN-SHE}
and references therein). In particular, the theoretical description of fission in superheavy elements is  quite relevant
to better understand both the shell structure evolution and  the appearance of new proton and/or neutron magic numbers 
in heavy nuclei
\cite{Sobiczewski,Mosel}. Superheavy elements are also 
produced during the r-process and their properties determine  
the upper end of the nucleosynthesis flow \cite{Arnould-2007}. 

%%%%%%%%%%%%%%%%%%%%%%%%%%%%%%%%%%%%%%%%%%%%%%%%%%%%%%%%%%%%%%%%%%%%%%%%%%%%%%%%%%%%%%%%%%%%%%%%%
%
%   FIGURE:   H F B       E N E R G I E S   I N    2 4 0   U 
%
%%%%%%%%%%%%%%%%%%%%%%%%%%%%%%%%%%%%%%%%%%%%%%%%%%%%%%%%%%%%%%%%%%%%%%%%%%%%%%%%%%%%%%%%%%%%%%%%%
\begin{figure}
\includegraphics[width=0.45\textwidth]{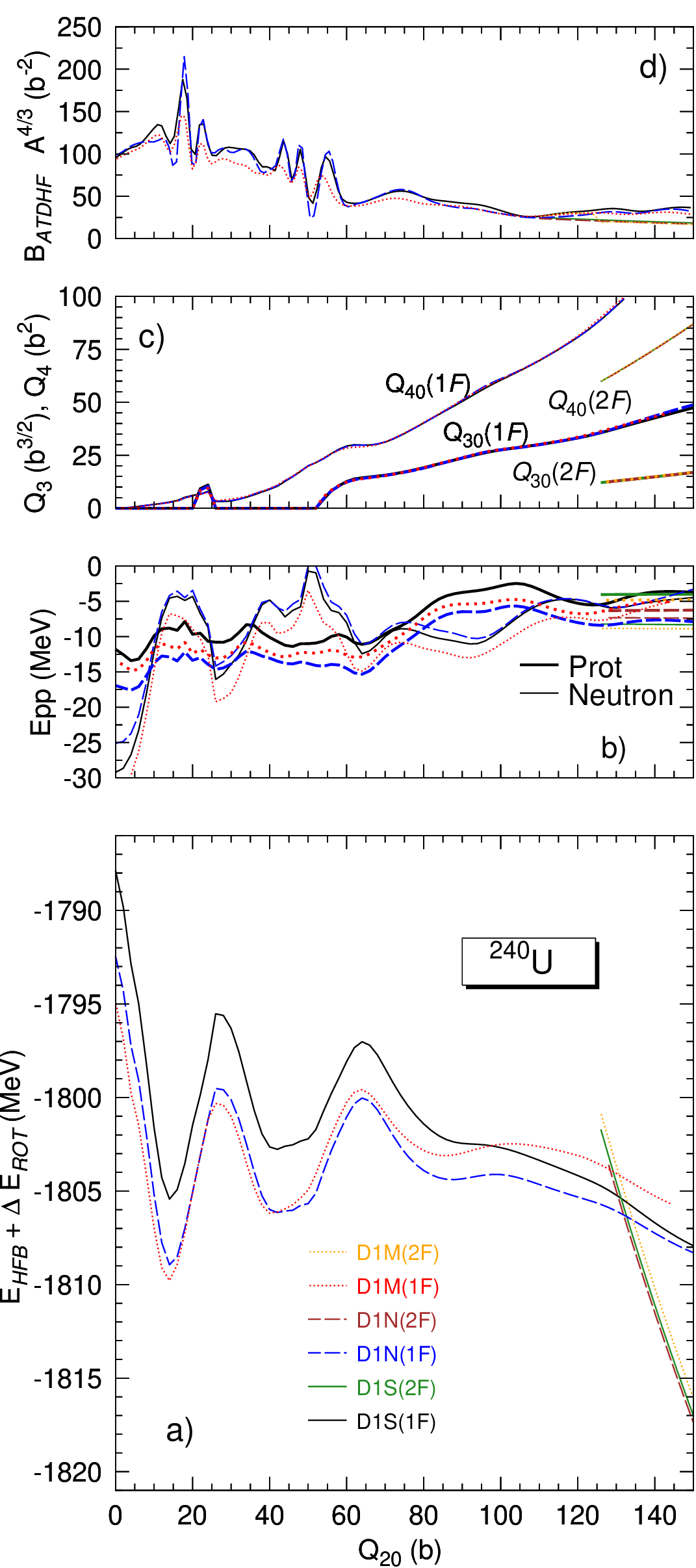}
\caption{ 
(Color online) The HFB plus the zero point rotational energies obtained 
with the D1S, D1N and D1M EDFs are plotted in panel a) as functions of the quadrupole 
moment $Q_{20}$ for the nucleus $^{240}$U. For each 
EDF, both the one (1F) and two-fragment (2F) solutions 
are included  in the plot. The pairing interaction energies are depicted 
in panel b) for protons (thick lines) and neutrons (thin lines). 
The octupole and hexadecapole moments 
corresponding to the 1F and 2F solutions are given in panel c). The collective masses
obtained within the ATDHFB approximation are plotted in panel d). For more
details, see the main text.
}
\label{FissionBarriersD1SD1ND1M240U} 
\end{figure}
%%%%%%%%%%%%%%%%%%%%%%%%%%%%%%%%%%%%%%%%%%%%%%%%%%%%%%%%%%%%%%%%%%%%%%%%%%%%%%%%%%%%%%%%%%%%%%%%%
 
In addition, it should be kept in mind that, as a decay mode, spontaneous fission 
competes with $\alpha$-decay \cite{Viola-Seaborg} and determines 
the stability of heavy and  superheavy elements. It is therefore, highly 
desirable to devote systematic microscopic studies, based on different 
effective EDFs, to the prediction of the  spontaneous fission $t_\mathrm{SF}$
and $\alpha$-decay $t_{\alpha}$ half-lives (see, for example, Refs. \cite{Warda-Egido-2012,Erler2012}). This is 
particularly  relevant, taking into account the large uncertainties \cite{Robledo-Giulliani} associated with 
the different building blocks entering the Wentzel-Kramers-Brillouin (WKB) formula 
\cite{Baran-TSF-1,Baran-TSF-2} used to computed the  $t_\mathrm{SF}$ values.  

Although the theoretical
uncertainties in the determination of the absolute values of the fission 
observables are presumed to be large \cite{Robledo-Giulliani},
the behavior of quantities as a function of mass number and/or 
along isotopic chains is reasonably well reproduced.
Therefore, one expects to obtain a reasonable theoretical 
description of the physics of fission along isotopic
chains extending up to the neutron dripline. Those regions are 
the territories where the fate of the nucleosynthesis of
heavy nuclei is determined. To study the fission of neutron-rich 
nuclei we have used a mean-field framework with the Gogny-EDF 
in the Uranium isotopic chain up to the neutron dripline nucleus
 $^{280}$U. The three most relevant parametrizations of 
the Gogny-EDF \cite{gogny}, namely D1S \cite{gogny-d1s}, D1N 
\cite{gogny-d1n} and D1M \cite{gogny-d1m} have been used in the 
calculations. The D1S parametrization is the oldest among the three 
and its fitting protocol included fission properties of $^{240}$Pu. 
Along the years, D1S has built itself a strong reputation given its 
ability to reproduce a large collection of low-energy data all over 
the periodic table \cite{gogny-d1s,gogny,PRCQ2Q3-2012,Warda-Egido-Robledo-Pomorski-2002,PTpaper-Rayner,
Delaroche-2006,RRG23S,ER-Lectures,REL-2008,Rayner-PRL2004,Rayner-PRC2004,Berger-1989,Chin-1992,
Egido-other1,Egido-other2,Egido-other3,Girod-1989,Gogny-Bertsch-2007,Peru-2005,NPA-2002,
Rayner-PRC-2002,Hilare-2007,Delaroche-2010,Warda-Egido-2012}. In particular, the 
parametrization D1S has already been successfully applied to the 
microscopic description of fission in heavy and superheavy nuclei
(see, for example, Refs. \cite{Warda-Egido-Robledo-Pomorski-2002,Delaroche-2006,Warda-Egido-2012} and 
references therein) and, for this reason, it is taken as a reference in the 
present study. However, D1S is not specially good in reproducing masses specially 
when moving away from the stability valley. To cure this deficiency the D1N parametrization
was introduced. It provides a good fit to realistic neutron matter equation of state 
(EoS)
and therefore it is
expected to perform well in dealing with neutron-rich nuclei. However, this Gogny-EDF has scarcely being
used and its performance in fission has to be validated. Finally, the D1M parametrization included in its
fitting protocol not only  realistic  neutron matter EoS information but also the binding energies of all known nuclei. 
With an impressive rms for the binding energy of 0.798 MeV it represents 
an excellent and competitive choice to deal with nuclear masses.
An extensive program to establish the merits and shortcomings of 
D1M in nuclear structure studies not only in even-even nuclei 
\cite{PTpaper-Rayner,gogny-d1n,gogny-d1m,Rayner-Sara,Rayner-Robledo-JPG-2009,Rayner-PRC-2010,
Rayner-PRC-2011,PRCQ2Q3-2012,Robledo-Rayner-JPG-2012}, 
but also in odd-A ones in the framework of the equal filling approximation
(EFA) \cite{Rayner-Sara,Rayner-PRC-2010,Rayner-PRC-2011} is in progress. 
However, this parametrization has not been used systematically in  fission studies before, 
and therefore its properties regarding fission
have to be validated as in the case of D1N. As a consequence of these needs, 
we have decided to carry out calculations with the D1M parametrization 
for a selected set of nuclei consisting of $^{238-244}$Pu, $^{240-248}$Cm,
$^{250,252}$Cf, $^{250-256}$Fm, $^{252-256}$No, $^{256-260}$Rf, 
$^{258-262}$Sg, $^{264}$Hs and the $Z=114$ nucleus $^{286}$Fl, for which experimental data are available  
\cite{Refs-barriers-other-nuclei-1,Refs-barriers-other-nuclei-2,Refs-barriers-other-nuclei-3-tsf,
Pu-mass-fragments-exp-1,Pu-mass-fragments-exp-2}. The comparison 
with these data and other quantities like fission barrier heights 
and excitation energies of fission isomers will be used to 
validate the results obtained with D1M. Later, calculations of fission properties for the Uranium chain 
from $^{232}$U up to the neutron dripline $^{280}$U
will be carried out. The comparison of the results obtained 
with the three parametrizations will serve to give
us an idea of the uncertainties associated to the Gogny-EDF used.

The paper is organized as follows. In Sec. \ref{Theory-used}, we briefly outline the theoretical
formalism used in the present work. The results
of our calculations are discussed in Sec. \ref{RESULTS}. First, in Sec. \ref{strategy-240U}, we illustrate
the methodology employed
to compute the fission paths and other fission-related quantities in the case of  $^{240}$U. The same 
methodology has been used for all the nuclei studied in this paper. In 
Sec. \ref{NucleiWithData}, we discuss the D1M results  
for the nuclei $^{232-238}$U, $^{238-244}$Pu, $^{240-248}$Cm,$^{250,252}$Cf, $^{250-256}$Fm, $^{252-256}$No, $^{256-260}$Rf, 
$^{258-262}$Sg, $^{264}$Hs and $^{286}$Fl and compare them with available experimental data
\cite{Refs-barriers-other-nuclei-1,Refs-barriers-other-nuclei-2,Refs-barriers-other-nuclei-3-tsf,Pu-mass-fragments-exp-1,Pu-mass-fragments-exp-2}. 
This section, is mainly  intended  to validate D1M for fission studies. The
systematics of the  fission paths, spontaneous fission half-lives and fragment mass 
in the  isotopes $^{232-280}$U is presented in Sec. \ref{FB-systematcis}.  We will 
compare the results obtained with the D1S, D1N and D1M parametrizations
to demonstrate the robustness of the predicted 
trends in $^{232-280}$U with respect to particular choices of parametrizations.
One of the main advantages of all the considered Gogny-EDFs is that they
provide a self-contained approach to pairing correlations  \cite{Gogny-1980}. 
Due to the differences in the corresponding fitting protocols
\cite{gogny-d1s,gogny-d1n,gogny-d1m}, each 
of the EDFs displays a different pairing content
\cite{PTpaper-Rayner}. This, by itself, provides some 
insight into the impact of pairing correlations on fission 
properties in $^{232-280}$U. However, in Sec. \ref{change-pairing-strenght}, we
explicitly discuss the impact of pairing correlations on the 
predicted $t_\mathrm{SF}$ values for $^{232-280}$U by increasing artificially 
the pairing strengths by 5 and 10 $\%$, respectively. Finally, Sec. \ref{conclusions}
is devoted to the concluding remarks and work perspectives.

% ----------------------------------------------------------------------
%
%  S E C T I O N
%
%          T h e o r e t i c a l     f r a m e w o r k 
%
% ----------------------------------------------------------------------  

\section{Theoretical framework}
\label{Theory-used}

The mean-field approximation \cite{rs} based on wave functions $| 
\Phi_{HFB} \rangle$ of the HFB type has 
been used in the present study. Constraints in the mean value of the 
axially symmetric quadrupole  $\hat{Q}_{20}$, octupole $\hat{Q}_{30}$
as well as the necking $\hat{Q}_{Neck}(z_{0},C_{0})$ operators have 
been used. The last constraint, as discussed in Sec. \ref{strategy-240U},
allows us to reach two-fragment (2F) solutions starting from the one-fragment (1F) ones 
\cite{Robledo-Giulliani,Warda-Egido-Robledo-Pomorski-2002,Egido-other1}.  
As a consequence of the axial symmetry imposed on our HFB wave 
functions $| \Phi_{HFB} \rangle$, the mean values of the multipole 
operators $\hat{Q}_{\mu \nu}$ with $\nu \ne 0$ are zero  by 
construction. Aside from the constraints already mentioned, as well 
as the usual  ones on both the proton and neutron numbers, a 
constraint on the operator $\hat{Q}_{10}$ is used  to prevent 
spurious effects associated to the center of mass motion.

The HFB quasiparticle operators \cite{rs} have been expanded in an 
axially symmetric (deformed)
harmonic oscillator  (HO) basis 
containing  states with $J_{z}$ quantum numbers up to 35/2 and up to
26 quanta in the z direction. The basis quantum numbers are restricted
by the condition
\begin{equation}
2 n_{\perp} + |m| + \frac{1}{q} n_{z} \le M_{z, \mathrm{MAX}}	
\end{equation}
with $M_{z, \mathrm{MAX}}=17$ and $q=1.5$. This choice is well suited 
for the elongated prolate shapes  typical of the fission process 
\cite{Robledo-Giulliani,Warda-Egido-Robledo-Pomorski-2002}. 
For each of the considered nuclei and each of the constrained configurations  
($Q_{20},Q_{30},Q_{Neck}, \dots$)
the two lengths  $b_{z}$ and $b_{\perp}$ characterizing the HO basis have 
been optimized so as to minimize the total HFB energy. With the choice of basis
size and the minimization of the energy with the oscillator lengths, the  
relative energies determining the dynamics of the fission process are well converged. 
For the solution of the HFB equations, an 
approximate second order gradient method  \cite{Robledo-Bertsch2OGM} has been used.
The method is very robust and the typical number of iterations to converge
is quite small (a few tens) as compared to other methods. In addition, the  
complexity in the handling of constraints does not increase with its number.

Concerning the different interaction terms, the two-body kinetic 
energy correction has been  fully taken into account (including 
exchange and pairing channels) in the variational  procedure. 
On the other hand, the Coulomb exchange term 
is considered in the Slater approximation \cite{CoulombSlater} while 
the Coulomb and spin-orbit contributions to the pairing field have 
been neglected.

%%%%%%%%%%%%%%%%%%%%%%%%%%%%%%%%%%%%%%%%%%%%%%%%%%%%%%%%%%%%%%%%%%%%%%%%%%%%%%%%%%%%%%%%%%%%%%%%%
%
%   FIGURE 2 OF THE PAPER  
%
%%%%%%%%%%%%%%%%%%%%%%%%%%%%%%%%%%%%%%%%%%%%%%%%%%%%%%%%%%%%%%%%%%%%%%%%%%%%%%%%%%%%%%%%%%%%%%%%%
\begin{figure}
\includegraphics[width=0.5\textwidth]{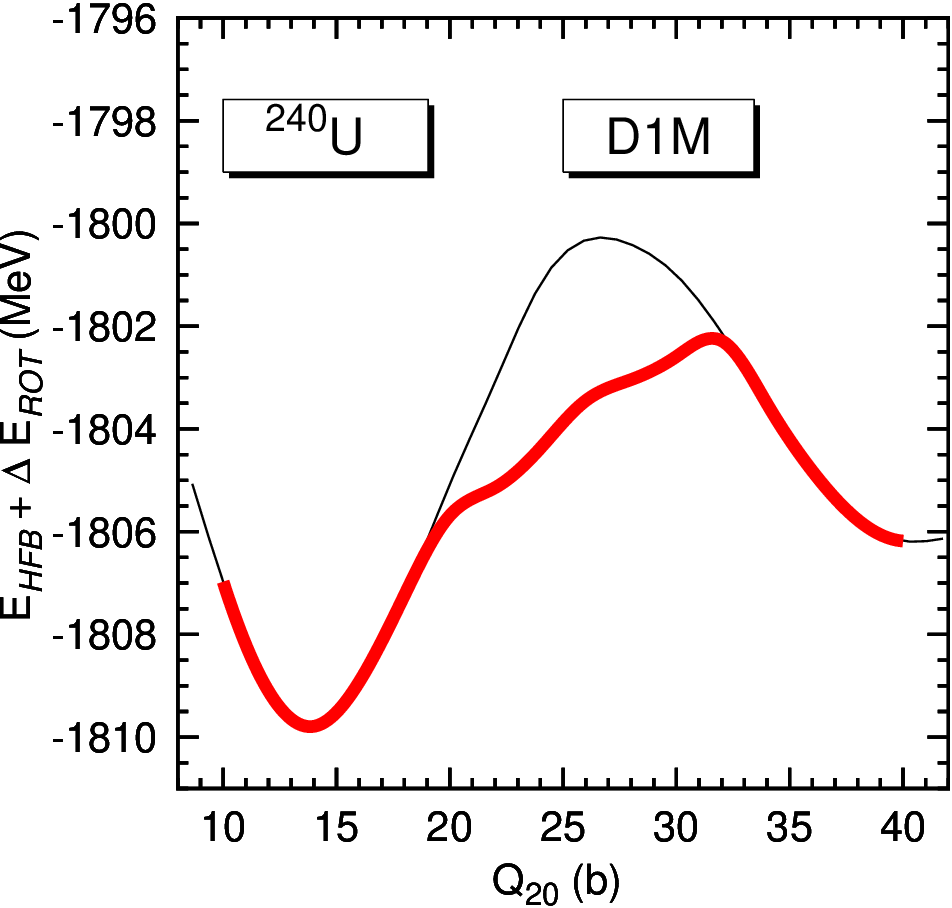}
\caption{ 
(Color online) The HFB plus the zero point rotational energies obtained in the 
framework of axially symmetric  calculations (black thin curve), based on the Gogny-D1M EDF, for
the nucleus $^{240}$U are compared with the ones provided by triaxial  calculations (red thick curve). 
Results are shown for configurations around the inner fission barrier 
(see, Fig. \ref{FissionBarriersD1SD1ND1M240U}).
}
\label{FissionBarrierD1M240Uplusgamma} 
\end{figure}
%%%%%%%%%%%%%%%%%%%%%%%%%%%%%%%%%%%%%%%%%%%%%%%%%%%%%%%%%%%%%%%%%%%%%%%%%%%%%%%%%%%%%%%%%%%%%%%%%

The spontaneous fission half-life is computed (in seconds) with 
the  WKB formalism \cite{proportional-1} as 
\begin{eqnarray} \label{TSF}
t_\mathrm{SF}= 2.86 \times 10^{-21} \times \left(1+ e^{2S} \right)
\end{eqnarray}
where the action S along the quadrupole constrained fission path reads
\begin{eqnarray} \label{Action}
S= \int_{a}^{b} dQ_{20} \sqrt{2B(Q_{20})\left(V(Q_{20})-\left(E_{GS}+E_{0} \right)  \right)}.
\end{eqnarray}  
Here the integration limits $a$ and $b$ are the classical turning points
\cite{proportional-1} below the barrier and corresponding to the
energy $E_{GS}+E_{0}$. The potential  $V(Q_{20})$ is given by the HFB energy 
corrected by the  
zero point energies  stemming from the restoration of the rotational symmetry 
$\Delta E_{ROT}(Q_{20})$ and 
the fluctuations in the quadrupole moment $\Delta E_{vib}(Q_{20})$. The  rotational
correction $\Delta E_{ROT}(Q_{20})$ has been computed, in terms of the Yoccoz moment of 
inertia, according to the phenomenological
prescription discussed in Refs. \cite{RRG23S,ER-Lectures}. This correction 
plays a key role to determine the shape of the potential $V(Q_{20})$ as  
it can be as large as 6 -- 7 MeV and its value is proportional to the degree
of symmetry breaking, i.e., the value of the deformation $Q_{20}$ \cite{NPA-2002}.   

For the evaluation of the collective mass $B(Q_{20})$ and the vibrational 
energy correction $\Delta E_{vib}(Q_{20})$ two methods have been used. One is the 
cranking approximation 
\cite{crankingAPPROX,Giannoni,Libert-1999}
to the Adiabatic Time Dependent HFB (ATDHFB) scheme \cite{rs}. In this case
\begin{eqnarray} \label{mass-ATDHFB}
B_{ATDHFB}(Q_{20})= \frac{1}{2} \frac{{\cal{M}}_{-3}(Q_{20})}{{\cal{M}}_{-1}^{2}(Q_{20})}
\end{eqnarray} 
where the moments ${\cal{M}}_{-n}(Q_{20})$ of the 
generating quadrupole field read
\begin{eqnarray} \label{M-def}
{\cal{M}}_{-n}(Q_{20}) = \sum_{\mu \nu} \frac{|\hat{Q}_{\mu \nu}^{20}|^{2}}{\left(E_{\mu} + E_{\nu} \right)^{n}}
\end{eqnarray}
and $\hat{Q}_{\mu \nu}^{20}$ is the $20$-component  of the quadrupole operator in the 
quasiparticle representation \cite{rs}. The quasiparticle energies $E_{\mu}$ are the ones 
obtained in the solution of the  HFB equations. The  ATDHFB zero point 
vibrational correction $\Delta E_{vib}(Q_{20})$ is  given by
\begin{eqnarray} \label{vibATDHFB}
\Delta E_{vib,ATDHFB}(Q_{20}) = \frac{1}{2} G(Q_{20}) B_{ATDHFB}^{-1}(Q_{20})
\end{eqnarray}
where 
\begin{eqnarray}
G(Q_{20}) = \frac{1}{2} \frac{{\cal{M}}_{-2}(Q_{20})}{{\cal{M}}_{-1}^{2}(Q_{20})}
\end{eqnarray}
is the width of the overlap between two configurations with similar quadrupole 
moments.

%%%%%%%%%%%%%%%%%%%%%%%%%%%%%%%%%%%%%%%%%%%%%%%%%%%%%%%%%%%%%%%%%%%%%%%%%%%%%%%%%%%%%%%%%%%%%%%%%
%
%   FIGURE 3 OF THE PAPER  
%
%%%%%%%%%%%%%%%%%%%%%%%%%%%%%%%%%%%%%%%%%%%%%%%%%%%%%%%%%%%%%%%%%%%%%%%%%%%%%%%%%%%%%%%%%%%%%%%%%
\begin{figure*}
\includegraphics[width=0.3\textwidth]{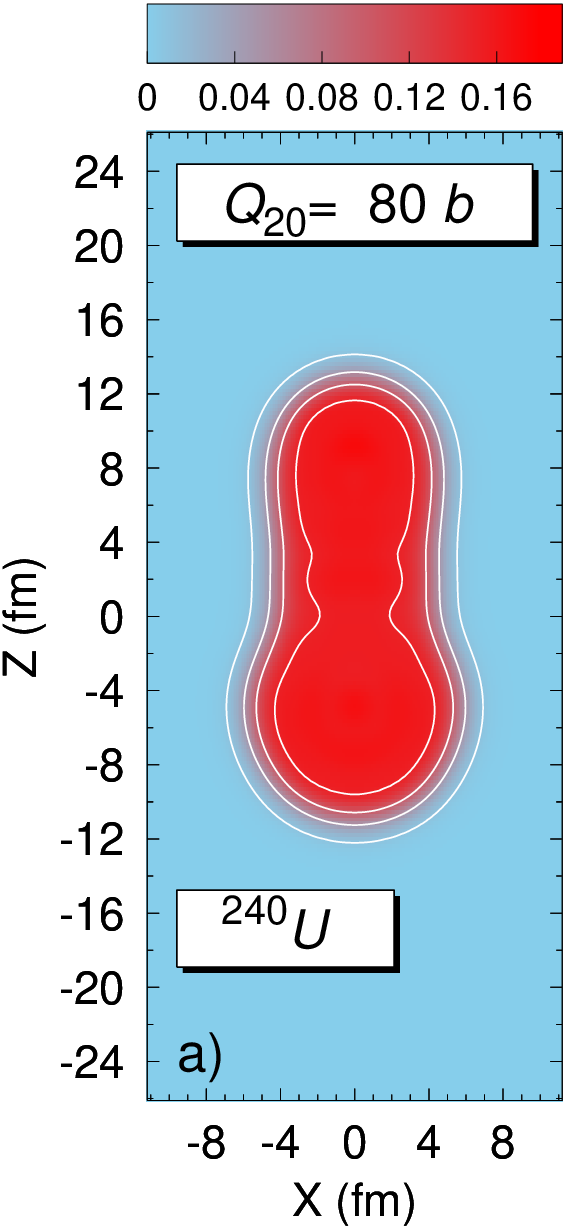}
\includegraphics[width=0.3\textwidth]{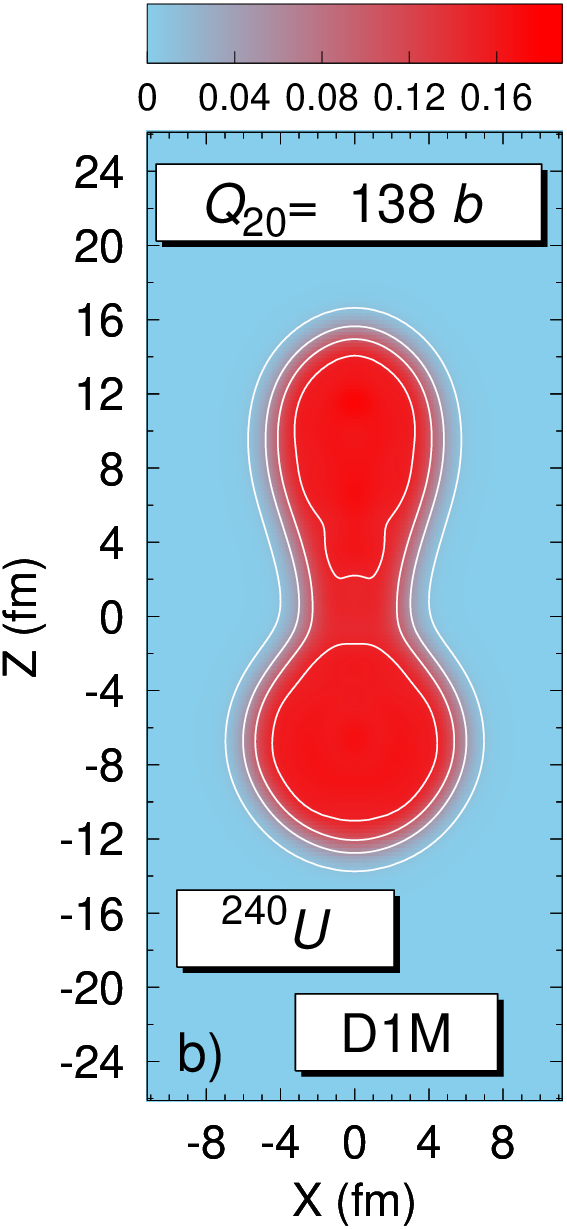}
\includegraphics[width=0.3\textwidth]{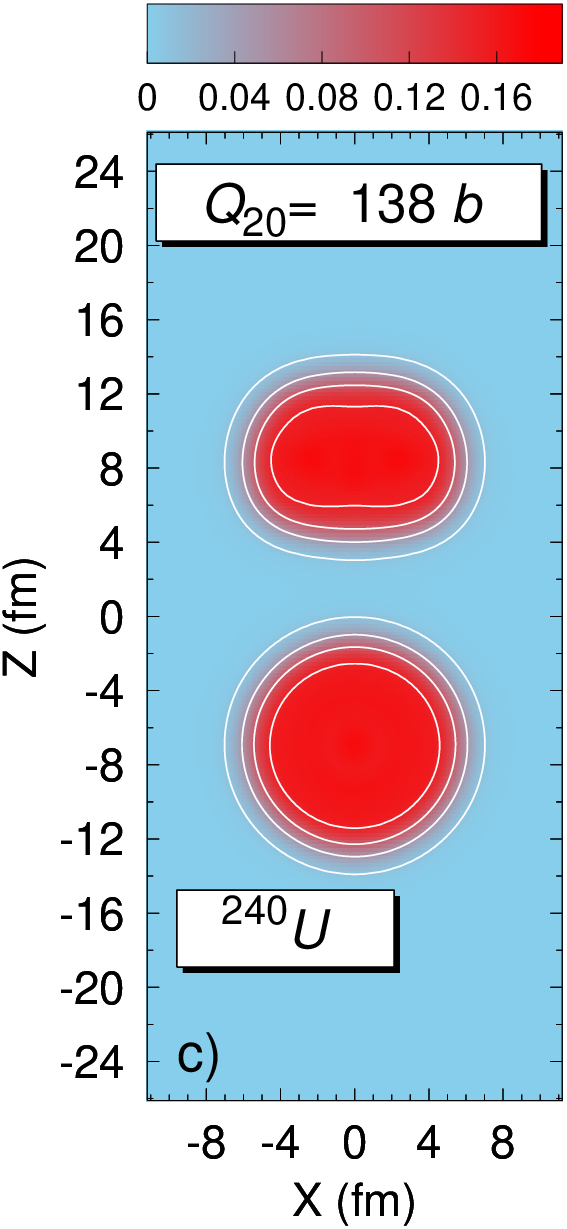}
\caption{ (Color online) Density contour plots for the nucleus $^{240}$U at the 
quadrupole deformations $Q_{20}$=80 b [panel a)] and  $Q_{20}$=138 b [panels b) and c)]. The
density profiles in panels a) and b) correspond to 1F configurations while the one 
in panel c) represents a 2F solution. Results are shown 
for the parametrization D1M of the Gogny-EDF. Densities are in units of fm$^{-3}$ and contour
lines at drawn at 0.01, 0.05, 0.10 and 0.15 fm$^{-3}$.
}
\label{den_cont_240U_D1M} 
\end{figure*}
%%%%%%%%%%%%%%%%%%%%%%%%%%%%%%%%%%%%%%%%%%%%%%%%%%%%%%%%%%%%%%%%%%%%%%%%%%%%%%%%%%%%%%%%%%%%%%%%%

The second method is based on the Gaussian Overlap Approximation (GOA)  to the 
GCM \cite{rs}. Here, the collective mass reads
\begin{eqnarray} \label{mass-GCM}
B_{GCM}(Q_{20})= \frac{1}{2} \frac{{\cal{M}}_{-2}^{2}(Q_{20})}{{\cal{M}}_{-1}^{3}(Q_{20})}
\end{eqnarray}
and $\Delta E_{vib,GCM}(Q_{20})$ is given by  Eq. (\ref{vibATDHFB}) but 
replacing the ATDHFB mass with the GCM one.
We have evaluated the spontaneous fission half-life $t_\mathrm{SF}$ Eq. (\ref{TSF}) with
the two schemes outlined above. The reason is 
that the ATDHFB masses are typically around 1.5 to 2  times larger
than the GCM ones  \cite{Robledo-Giulliani,Baran-mass-2011}. 
As a consequence, the action in the exponent defining $t_\mathrm{SF}$ is, 
in the ATDHFB case, between  20  and  40 \% larger  than the GCM one.
 Depending on the value of the action, this increase can represent a difference of several orders of 
 magnitude in the $t_\mathrm{SF}$ results. We also have to keep in
 mind that the inertias are computed in the so called "perturbative cranking approximation"
 that is known to underestimate the real inertia values by a factor as small as
 0.7 implying a reduction of a typical 15 \% in the action.
 For a thorough comparison of different  forms of the collective
 inertia in the framework of Skyrme-like EDFs, including the ones in 
 Eqs. (\ref{mass-ATDHFB}) and (\ref{mass-GCM}), and including also 
 the different computational schemes the 
 reader is  referred to Ref. \cite{Baran-mass-2011}.

In Eq. (\ref{Action}), the parameter $E_{0}$  
accounts for the true ground state energy once the 
zero point quadrupole fluctuations are considered. Although it is
not difficult to estimate its value using the curvature of the energy
around the ground state minimum and the values of the collective inertias 
\cite{Baran-SF-2012} we have followed the usual 
recipe \cite{Robledo-Giulliani,Warda-Egido-Robledo-Pomorski-2002}
of considering it as a free parameter that takes four different values
(i.e., $E_{0}$=0.5, 1.0, 1.5 and 2.0 MeV). In this way we can estimate 
its impact  on the predicted  spontaneous fission half-lives. 

To summarize the previous discussions, we conclude that 
the $t_\mathrm{SF}$ values obtained within our computational scheme are subject to
several uncertainties related to the following items :

\begin{enumerate}

\item The characteristics of the different parametrizations of the Gogny-EDF considered. 

\item The impact of triaxiality in the fission path. It is well known
that the configurations around the top of the inner barrier can reduce their
energies  when triaxiality is allowed. It is also 
possible in some superheavy nuclei that their oblate ground state evolves
towards fission through a triaxial path. In our case,  we have
kept axial symmetry as a self-consistent symmetry along the whole fission 
path in order to reduce the already substantial computational effort. However,
for a few selected configurations around the inner barrier we have allowed
triaxiality to set in as to study the reduction of the inner barrier height.
Typically, the lowering represents at most a few MeV  when triaxial 
shapes are allowed  \cite{Abusara-2010,Delaroche-2006}. However, the
 lowering of the inner barrier comes together
with an increase of the  collective inertia \cite{Bender-1998,Baran-1981} that
tends to compensate in the final value of the action. Therefore, the impact
of triaxiality in the final value of $t_\mathrm{SF}$ is very limited and 
it has not been considered in the present study. In addition, 
previous studies \cite{Baran-1981} analyzing the {\it{dynamical}} path to 
fission have corroborated the insignificant role played by triaxiality to
determine lifetimes. 

\item The value of the parameter $E_{0}$. This is particularly relevant 
in the case of long-lived isotopes with wide and high fission barriers since the 
different $E_{0}$ values provide different classical turning points $a$ and 
$b$ [see, Eq. (\ref{Action})] and therefore modify in a substantial way the
final value of the action integral.

\item The assumptions involved in the computation  of the
collective masses as well as  the zero point corrections
to the HFB energies. Note that, for example, within the `` perturbative 
cranking"  scheme \cite{crankingAPPROX,Giannoni,Libert-1999}, only the 
zero-order approximation is used instead of the  full linear-response 
matrix.

\item  Pairing correlations. They play a key role in the computation of 
both the zero point energies associated to quantum fluctuations 
and the collective masses. In fact, as we will see in Sec. \ref{change-pairing-strenght}
(see also Ref. \cite{Robledo-Giulliani}), changes of 5 or 10 $\%$
in the pairing strengths of the original Gogny-D1M EDF
can modify the predicted  $t_\mathrm{SF}$ values by several orders
of magnitude.

\end{enumerate}

As a consequence the predicted $t_\mathrm{SF}$ values will have large theoretical
error bars spanning several orders of magnitude implying that their absolute
values cannot be used with confidence. However, the experimental isotopic and/or isotonic
trends are reproduced with much higher accuracy giving us confidence
on the validity of our predictions in that respect.

Finally, we have  computed the $\alpha$-decay half-lives using the 
parametrization \cite{TDong2005} 
\begin{eqnarray} \label{VSeaborg-new}
\log_{10} t_{\alpha} =  \frac{AZ+B}{\sqrt{ {\cal{Q}}_{\alpha}}} + CZ+D
\end{eqnarray}
of the phenomenological Viola-Seaborg formula \cite{Viola-Seaborg}. 
The ${\cal{Q}}_{\alpha}$ value (in MeV) is obtained from the calculated 
binding energies for  Uranium  and Thorium isotopes as 
\begin{eqnarray} \label{QALPHA}
{\cal{Q}}_{\alpha}= E(Z,N) -E(Z-2,N-2) - E(2,2)
\end{eqnarray} 
In Eqs. (\ref{VSeaborg-new}) and (\ref{QALPHA}), $Z$ and $N$ 
represent the proton and neutron  numbers of the parent nucleus. On the other 
hand, $E(2,2)$=-28.295674 MeV \cite{AudiWapstraThibault-2003} while 
$A$=1.64062, $B$=-8.54399, $C$=-0.19430 and $D$=-33.9054 \cite{TDong2005}. 	

% ----------------------------------------------------------------------
%
%                                                       T A B L E     I 
% 
% ----------------------------------------------------------------------

\begin{table}
\label{Table1}
\caption{
The heights of the inner  $B_{I}^\mathrm{th}$ and second $B_{II}^\mathrm{th}$
barriers as well as the excitation energies $E_{II}^\mathrm{th}$ of the fission 
isomers, predicted with the Gogny-D1M EDF, are compared with the available 
experimental values  $B_{I}^\mathrm{exp}$, $B_{II}^\mathrm{exp}$ and $E_{II}^\mathrm{exp}$ 
\cite{Refs-barriers-other-nuclei-1,Refs-barriers-other-nuclei-2}. 
The  $B_{I}^\mathrm{th}$ values obtained in the framework 
of triaxial calculations are given in parenthesis. All the theoretical
results have been obtained from the rotational corrected HFB  energies. 
For more details, see the main text.
}
\begin{tabular}{ccccccc}
\hline\hline
Nucleus   &  $B_{I}^\mathrm{th}$  &  $B_{I}^\mathrm{exp}$ & $E_{II}^\mathrm{th}$  &  $E_{II}^\mathrm{exp}$ & $B_{II}^\mathrm{th}$  &  $B_{II}^\mathrm{exp}$  \\
\hline\hline
$^{234}$U &   7.60           &  4.80          &  3.32          &    -            &  8.09          &  5.50            \\ 
          &  (7.01)          &                &                &                 &                &                  \\
      
$^{236}$U &  8.33            &  5.00          &  3.17          &   2.75          &  8.69          &  5.67              \\
          & (7.00)           &                &                &                 &                &                  \\
         
$^{238}$U &  9.06            &  6.30          &  3.37          &   2.55          &  9.54          &  5.50                 \\
          & (7.46)           &                &                &                 &                &                  \\
      
$^{238}$Pu &  8.77           &  5.60          &  3.20          &   2.40          &  7.75          &  5.10                 \\
           & (7.66)          &                &                &                 &                &                  \\
       
$^{240}$Pu &  9.45           &  6.05          &  3.36          &   2.80          &  8.57          &  5.15                 \\
           & (7.70)          &                &                &                 &                &                  \\
         
$^{242}$Pu &  9.90           &  5.85          & 3.57           &   2.20          &  9.18          &  5.05                 \\
           & (7.67)          &                &                &                 &                &                  \\
          
$^{244}$Pu &  10.16          &  5.70          & 3.83           &   -             &  9.60          &  4.85                 \\
           & (7.42)          &                &                &                 &                &                  \\
         
$^{240}$Cm &  8.98           &  -             & 2.55           &   2.00          & 6.13           &  -                 \\
           & (7.87)          &                &                &                 &                &                  \\
        
$^{242}$Cm &  9.78           &  6.65          & 2.77           &   1.90          & 6.99           &  5.00                 \\
           & (8.31)          &                &                &                 &                &                  \\
         
$^{244}$Cm & 10.38           &  6.18          & 3.02           &   2.20          & 7.70           &  5.10                 \\
           & (8.27)          &                &                &                 &                &                  \\
         
$^{246}$Cm &  10.75          &  6.00          & 3.29           &   -             & 8.13           &  4.80                 \\
           & (8.03)          &                &                &                 &                &                  \\
          
$^{248}$Cm & 10.68           &  5.80          & 3.32           &   -             & 8.28           &  4.80                 \\
           & (7.50)          &                &                &                 &                &                  \\
         
$^{250}$Cf &  11.38          &  -             & 2.81           &   -             & 7.09           &  3.80                 \\
           & (8.25)          &                &                &                 &                &                  \\
          
$^{252}$Cf &  10.96          &  -             & 1.37           &   -             & 6.79           &  3.50                 \\
           & (8.07)          &                &                &                 &                &                  \\
\hline\hline
\end{tabular}
\end{table}
% ------------------------------------------------------------------------------------------------------

% ----------------------------------------------------------------------
%
%  S E C T I O N
%
%         D i s c u s s i o n    o f    t h e    r e s u l t s  
%
% ----------------------------------------------------------------------  

\section{Discussion of the results}\label{RESULTS}
In this section, we discuss all the results obtained. First, 
in Sec. \ref{strategy-240U}, we illustrate the methodology used
to compute the fission observables in the case of $^{240}$U. In 
Sec. \ref{NucleiWithData}, we discuss the  Gogny-D1M results 
for a  set of U, Pu, Cm, Cf, Fm, No, Rf, Sg and Fl nuclei for which 
experimental data are 
available \cite{Refs-barriers-other-nuclei-1,Refs-barriers-other-nuclei-2,Refs-barriers-other-nuclei-3-tsf,Pu-mass-fragments-exp-1,Pu-mass-fragments-exp-2}. 
The aim of these calculations is to validate D1M as a 
reasonable
parameter set for
fission studies. The systematics, provided by the D1S, D1N and D1M  
Gogny-EDFs, for the  fission paths, $t_\mathrm{SF}$ and 
$t_{\alpha}$ values as well as the fragment mass in the Uranium chain  
$^{232-280}$U is presented in Sec. \ref{FB-systematcis}. Finally, 
in Sec. \ref{change-pairing-strenght}, we explicitly discuss the impact 
of pairing correlations on the predicted $t_\mathrm{SF}$ values 
for $^{232-280}$U using a modified Gogny-D1M EDF in  which the pairing 
interaction strengths are increased by 5  and 10 $\%$, respectively.

% ----------------------------------------------------------------------
%
%  S U B S E C T I O N
%
%        T h e   n u c l e u s   2 4 0   Pu  
%
% ----------------------------------------------------------------------  

\subsection{An illustrative example: the nucleus $^{240}$U}
\label{strategy-240U}

In Fig. \ref{FissionBarriersD1SD1ND1M240U} (a), the evolution of the 
energy as a function of the mass quadrupole moment for the nucleus 
$^{240}$U as the system evolves from its ground state to very 
elongated shapes is shown. The results obtained with the D1S, D1N 
and D1M parametrizations are depicted. The energies shown in the 
plot are the HFB energies plus the ones coming from the zero point 
rotational motion $E_{HFB}+ \Delta E_{ROT}$. The zero point 
vibrational energies $\Delta E_{vib}$ (not included in the plot) are 
always considered in the evaluation of the lifetimes. The curves 
labeled D1S(1F) , D1N(1F) and D1M(1F), respectively, correspond to 
1F solutions of the HFB equations. In order to obtain such 1F 
solutions, we have first carried out reflection-symmetric $Q_{20}$
-constrained calculations. Subsequently, for each value of the 
quadrupole moment the octupole degree of freedom has been explored 
by constraining on a large value of $Q_{30}$ and then releasing the 
constraint to reach the lowest energy solution. Note, that 
constraints with higher multipolarities are not explicitly included 
in these calculations but, as it corresponds to a self-consistent 
calculation, the density profile (and therefore the mean value of 
the multipole moments) is determined as to minimize the energy. The 
only drawback of this procedure is that with our representation of 
the energy it is a mere  projection of a multidimensional path. As a 
consequence, kinks and multiple branches are common in this type of 
representation (the 2F quasifusion solution is an example). To help 
interpret the multidimensional energy surface, the values of $Q_{30}$
and $Q_{40}$ (to be discussed below) are very helpful.  

Coming back to  the figure, the three Gogny-EDFs provide 1F curves 
with rather similar shapes. The  ground state is located at $Q_{20}$
=14 b while  a first fission isomer appears at $Q_{20}$=42 b. Using 
the energies $E_{HFB}+ \Delta E_{ROT}$, we have obtained  (without 
triaxiality) the inner barrier heights 9.90, 9.42 and 9.47 MeV with 
the D1S, D1N and D1M parametrizations, respectively. Those values, 
for D1S and D1N, are in agreement with the values 20 and 18.4 MeV, 
respectively \cite{gogny-d1n} of the surface energy coefficient in 
nuclear matter $a_s$. In a semi classical picture of fission, the 
energy as a function of the driving coordinate (elongation) is the 
result of the competition between the increasing surface energy 
(governed by the $a_{s}$ value) and the decreasing Coulomb repulsion 
which is independent of the nuclear interaction. Unfortunately, the 
value of $a_{s}$ is not available in the literature for D1M. The 
excitation energies of the first fission isomers are 2.66, 2.79 and 
3.61 MeV for D1S, D1N and D1M, respectively. This quantity is 
associated to shell effects and it is usually believed to be 
strongly correlated with pairing correlations. Another noticeable 
feature from Fig. \ref{FissionBarriersD1SD1ND1M240U} (a) is the 
emergence of a second fission isomer around $Q_{20}$=86 b with its 
associated third fission barrier. As will be discussed later on in 
Sec. \ref{FB-systematcis}, such second fission isomers are also 
found in the 1F curves of several Uranium isotopes regardless of the 
Gogny-EDF employed \cite{Delaroche-2006}. Coming back to the second 
fission barrier, its height takes the values 8.41, 8.91 and 10.21 
MeV, for D1S, D1N and D1M, respectively. In this case, the trend 
observed in Ref \cite{Berger-1989} relating the height of the second 
barrier with $a_{s}$ (larger $a_{s}$ leads to larger barrier 
heights) is not fulfilled. A possible explanation is that at such 
large elongation the exchange properties of the interactions are 
more relevant than the surface properties. For the largest values of 
the quadrupole moment, the D1S and D1N curves show a similar decline 
due to the decreasing of Coulomb repulsion. In this region the D1S 
curve is a couple of MeV lower in energy than D1N. This is not 
consistent with the behavior observed in \cite{Berger-1989} for D1 
and D1S and attributed there to the $a_{s}$ values of the two 
interactions. For D1, with an $a_{s}$ coefficient 1.2 MeV larger 
than D1S, the HFB energy was around 10 MeV higher than for D1S. 
Finally, D1M shows a gentler decline than the ones provided by the 
D1N and D1S functionals. This points to a larger value of $a_{s}$ 
than for D1N and D1S but the first barrier height values point in 
the opposite direction of a lower surface energy coefficient for 
D1M. These results do not follow the neat trend observed in 
\cite{gogny-d1s,Berger-1989} in the comparison between the D1 and D1S 
parametrizations. This problem deserves further study, although a 
possible explanation is that the properties of the region beyond the 
first barrier are driven by quantum effects (exchange and shell 
effects) rather than macroscopic properties like the surface energy 
coefficient $a_{s}$.

%%%%%%%%%%%%%%%%%%%%%%%%%%%%%%%%%%%%%%%%%%%%%%%%%%%%%%%%%%%%%%%%%%%%%%%%%%%%%%%%%%%%%%%%%%%%%%%%%
%
%   FIGURE OF THE PAPER  
%
%%%%%%%%%%%%%%%%%%%%%%%%%%%%%%%%%%%%%%%%%%%%%%%%%%%%%%%%%%%%%%%%%%%%%%%%%%%%%%%%%%%%%%%%%%%%%%%%%
\begin{figure*}
\includegraphics[width=1.0\textwidth]{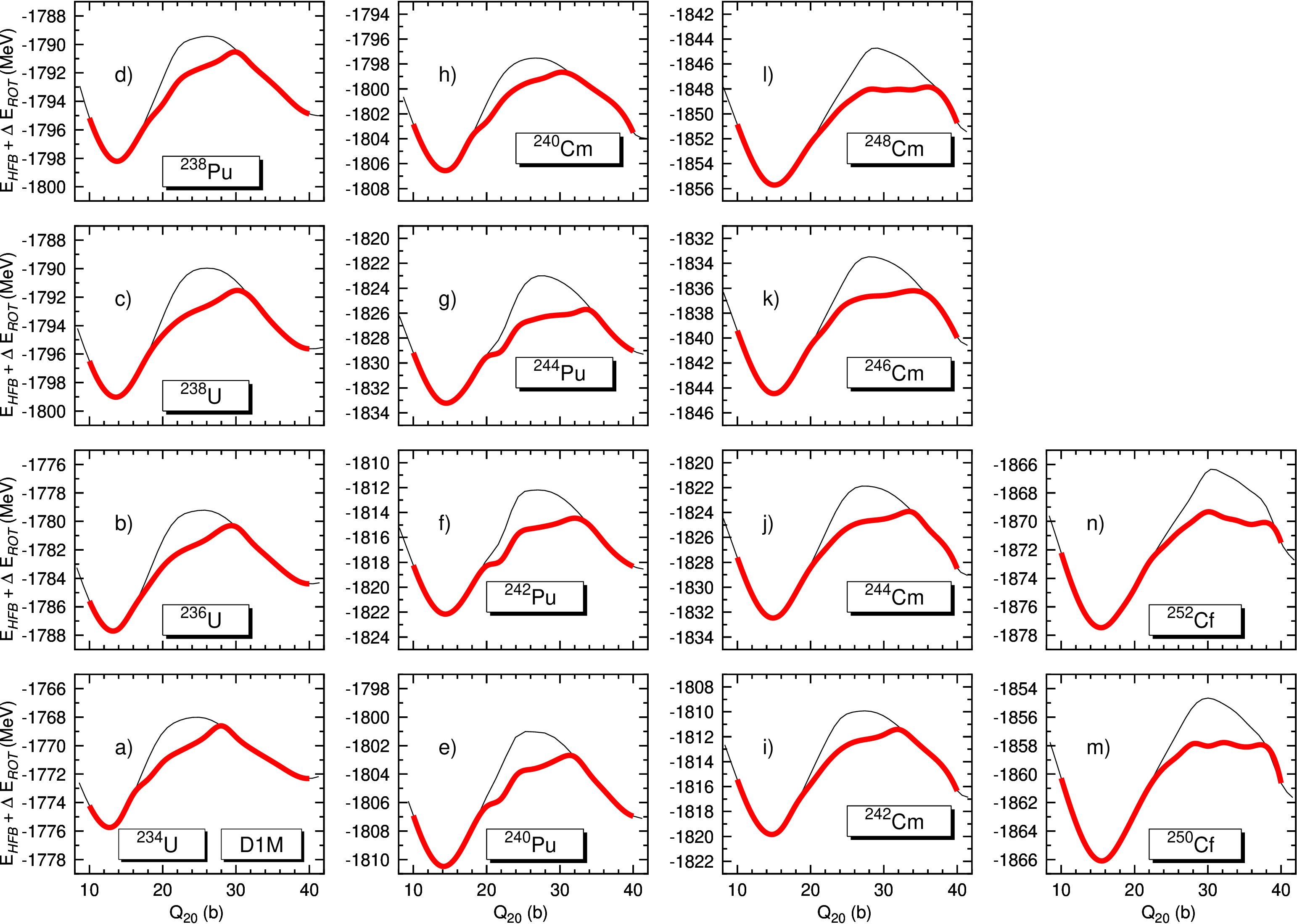}
\caption{ (Color online)
The HFB plus the zero point rotational energies obtained in the 
framework of axially symmetric  calculations (black thin curve), 
based on the Gogny-D1M EDF, for the nuclei $^{234-238}$U, 
$^{238-244}$Pu, $^{240-246}$Cm and $^{250,252}$Cf are compared with 
the ones provided by triaxial  calculations (red thick curve). 
Results are shown for configurations around the inner fission barrier.
}
\label{Heavy-Nuclei-gamma} 
\end{figure*}
%%%%%%%%%%%%%%%%%%%%%%%%%%%%%%%%%%%%%%%%%%%%%%%%%%%%%%%%%%%%%%%%%%%%%%%%%%%%%%%%%%%%%%%%%%%%%%%%%

In Fig. \ref{FissionBarrierD1M240Uplusgamma}, the 1F $E_{HFB}+ 
\Delta E_{ROT}$ energies obtained in the axially symmetric  
calculations  for the nucleus $^{240}$U are compared with the ones 
obtained in the framework of triaxial  calculations (see 
\cite{PTpaper-Rayner} for a thorough discussion of the framework and 
results with D1M). The curves depicted correspond to the D1M 
parametrization only, but similar results are obtained for the other 
parametrizations. The inclusion of the $\gamma$ degree of freedom 
leads to the  lowering of the energies in the  18 b $\le$ $Q_{20}$ 
$\le$ 32 b range   with $\gamma$ = 12$^{o}$ being the largest value 
in the region. Compared with the axially symmetric one (i.e., 9.47 
MeV) the height 7.51 MeV of the triaxial inner barrier in $^{240}$U 
displays a reduction of 1.96 MeV.
 
Coming back to  Fig. \ref{FissionBarriersD1SD1ND1M240U} (a), very 
steep curves labeled D1S(2F), D1N(2F) and D1M(2F) are depicted. 
They  correspond to  solutions with two well separated fragments and 
their energy corresponds to the quasifusion channel for the 
corresponding fragments. These 2F solutions can be reached, starting 
from the 1F ones, by constraining the hexadecapole moment 
$\hat{Q}_{40}$ \cite{gogny-d1s,Egido-other1}. Alternatively, one can 
resort to a constraint in the mean value of  the necking operator 
$\hat{Q}_{Neck}(z_{0},C_{0})$ \cite{Warda-Egido-Robledo-Pomorski-2002,Robledo-Giulliani}. 
For the  nucleus $^{240}$U, the 2F curves seem to intersect the 1F ones 
around $Q_{20}$=130 b and exhibit a quasilinear decrease in energy 
for increasing values of the quadrupole moment \cite{Robledo-Giulliani}. 
The intersection of the 1F and 2F curves, 
appears as  a consequence of projecting multi-dimensional paths into 
a one-dimensional plot. Actually, there is a minimum action path 
with a ridge connecting the 1F and 2F curves in the collective  
space. As the determination of this path is quite cumbersome and its 
contribution to the action Eq. (\ref{Action}) is small, we have 
neglected its contribution to the action. Within this approximation 
we take the 2F curves, for which the charge and mass of the 
fragments lead to the minimum energy, as really intersecting the 1F 
ones. In practice, we have obtained the 2F curves by constraining 
the number of particles in the neck of the parent nucleus to a small 
value and then releasing  the constraint in a  self-consistent HFB 
calculation. To asses the stability of the procedure a set of 
calculations with  different values of the neck parameters $z_{0}$ 
and $C_{0}$ \cite{Robledo-Giulliani} is performed to make sure that 
the same minimum is always reached.  The steep decrease in the 
energy of the 2F solutions is a consequence to the direct 
relationship that exists in this case between $Q_{20}$ and the 
fragments'  separation distance $R$. As the quadrupole moment of a 
2F solution increases the shape of the fragments remains more or 
less the same but the distance $R$ between them increases, 
decreasing the Coulomb repulsion between fragments and leading to 
the observed decrease of the energy \cite{gogny-d1s,Warda.11}. 

%%%%%%%%%%%%%%%%%%%%%%%%%%%%%%%%%%%%%%%%%%%%%%%%%%%%%%%%%%%%%%%%%%%%%%%%%%%%%%%%%%%%%%%%%%%%%%%%%
%
%   FIGURE OF THE PAPER  
%
%%%%%%%%%%%%%%%%%%%%%%%%%%%%%%%%%%%%%%%%%%%%%%%%%%%%%%%%%%%%%%%%%%%%%%%%%%%%%%%%%%%%%%%%%%%%%%%%%
\begin{figure}
\includegraphics[width=0.5\textwidth]{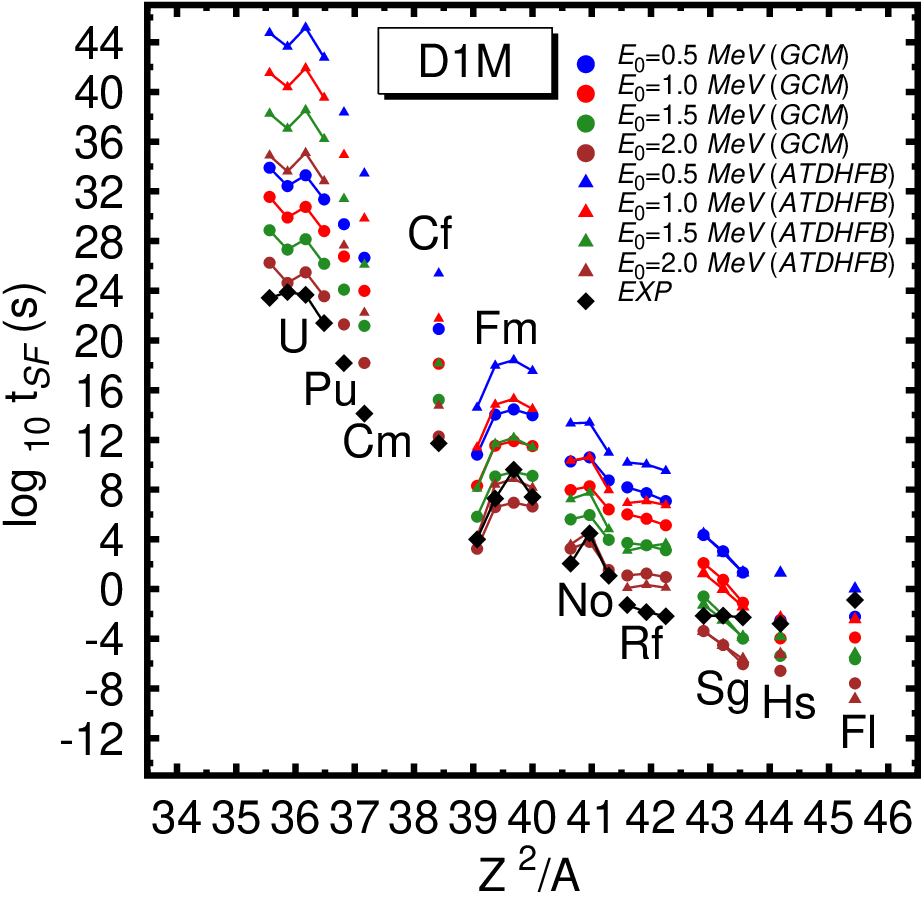}
\caption{ 
(Color online) The spontaneous fission half-lives $t_\mathrm{SF}$ obtained for the nuclei
$^{232-238}$U, $^{240}$Pu, $^{248}$Cm, $^{250}$Cf, $^{250-256}$Fm, $^{252-256}$No, $^{256-260}$Rf, 
$^{258-262}$Sg, $^{264}$Hs and $^{286}$Fl within the GCM and ATDHFB schemes are 
depicted as functions of the fissibility parameter $Z^{2}/A$ and compared with 
the corresponding experimental data \cite{Refs-barriers-other-nuclei-3-tsf}. 
The theoretical results, based on the Gogny-D1M EDF, are shown for 
$E_{0}$=0.5, 1.0, 1.5 and 2.0 MeV, respectively. For more details, see the main text.
}
\label{tsf-other-nuclei} 
\end{figure}
%%%%%%%%%%%%%%%%%%%%%%%%%%%%%%%%%%%%%%%%%%%%%%%%%%%%%%%%%%%%%%%%%%%%%%%%%%%%%%%%%%%%%%%%%%%%%%%%%

The proton  and neutron pairing interaction energies $E_{pp}=-1/2 
\mathrm{Tr}\left(\Delta \kappa^{*} \right)$, are shown in  Fig. \ref{FissionBarriersD1SD1ND1M240U} 
(b). In the three cases, they follow 
the same trend as functions of the quadrupole moment, being smaller 
for neutrons (protons) with the D1M (D1N) parametrization. In all 
cases, the neutron pairing energies display minima at $Q_{20}$=0 b, 
around the top of the inner and second fission barriers as well as 
around the second fission isomer. On the other hand, $E_{pp}$ 
exhibits maxima around $Q_{20}$=18 and 50 b, respectively. 

The  octupole and  hexadecapole moments are depicted, as functions of $Q_{20}$, in 
Fig. \ref{FissionBarriersD1SD1ND1M240U} (c). The 
values obtained with the three Gogny-EDFs can hardly be distinguished from each other. On the 
other hand, it is also
apparent from the figure, that the  moments corresponding to  the 1F 
[i.e., $Q_{30}(1F)$ and $Q_{40}(1F)$]  and 2F [i.e., $Q_{30}(2F)$ and $Q_{40}(2F)$] curves
are quite different reflecting, the separation of the paths in the multidimensional
space of parameters. 

The collective masses $B_{ATDHFB}$ 
are plotted in Fig. \ref{FissionBarriersD1SD1ND1M240U} (d). Their 
evolution, as functions of $Q_{20}$, is  well correlated with the one of the pairing 
interaction energies shown in Fig. \ref{FissionBarriersD1SD1ND1M240U} (b).
A similar pattern is found for the  GCM masses 
(not shown in the plot) though their 
values are always smaller than the ATDHFB ones. For example, for $Q_{20}$=18 b we have obtained 
the ratios $B_{ATDHFB}/B_{GCM}$= 1.97, 1.98 and 1.85 for the D1S, D1N and 
D1M parametrizations, respectively. 

With all the previous ingredients at hand, we have computed the spontaneous
fission half-lives using Eq. (\ref{TSF}). Since 
we take the 1F and 2F curves as intersecting ones and do not include 
the effect of triaxiality on the inner barriers, our $t_\mathrm{SF}$ values
should be regarded as lower bounds \cite{Robledo-Giulliani} to the real values. For the nucleus $^{240}$U, we have obtained
($E_{0}$ = 1.0 MeV) $t_\mathrm{SF}$ = 2.612 $\times$ 10$^{27}$ s, 2.161 $\times$ 10$^{35}$ s and 3.215 $\times$ 10$^{42}$ s
in the framework of the ATDHFB scheme for the D1S, D1N and D1M parametrizations, respectively. 
The large differences in the predicted fission half-lives 
can be  attributed to the differences in the fission paths  and 
ATDHFB masses provided by the considered EDFs. To disentangle the
different contributions we have taken the D1S fission path
and the D1N (D1M) ATDHFB mass to obtain 6.1 $\times 10^{26}$ s (1.5 $\times 10^{24}$ s) 
instead of the nominal value 2.612 $\times$ 10$^{27}$ s. We conclude that the main
effect is to be attributed to the different fission paths. The impact of the wiggles in
the masses has also been estimated by replacing the mass by a smoothed out 
mass (using a three point filter) and the half life changes by a factor 1.2
which is irrelevant in the present context. Using the GCM inertias, 
we obtain (again, $E_{0}$ = 1.0 MeV) smaller values 
$t_\mathrm{SF}$ = 4.089 $\times$ 10$^{20}$, 3.764 $\times$ 10$^{26}$ and 3.552 $\times$ 10$^{32}$ s.
Larger ATDHFB $t_\mathrm{SF}$ values as compared with the GCM ones is, as discussed
 in Sec. \ref{FB-systematcis}, a general trend for all the studied Uranium 
isotopes regardless of the 
particular functional employed. We thus see, how the differences in the ATDHFB and 
GCM masses, can have a strong impact on our predictions for fission observables.
This is the reason why, for each Gogny-EDF, both kinds of collective masses have been 
considered in the computation of the spontaneous fission half-lives. On the other hand, increasing
$E_{0}$ leads to smaller $t_\mathrm{SF}$ values in either the ATDHFB or GCM frameworks (see below for
a more quantitative assessment of the effect).

Finally the density contour plots corresponding to the nucleus $^{240}$U at the 
quadrupole deformations $Q_{20}$=80 and 138 b are shown in Figs. 
\ref{den_cont_240U_D1M} (a), (b) and (c). Results are shown only for the Gogny-D1M EDF but similar ones
have been obtained for the other parametrizations. For $Q_{20}$=138 b two plots are presented
corresponding to 1F and 2F solutions, respectively. The 2F solution in Fig. \ref{den_cont_240U_D1M} (c) consists
of a spherical $^{132}$Sn fragment plus  an oblate and slightly octupole  deformed $^{108}$Mo fragment with 
$\beta_{2}$=-0.22 and $\beta_{3}$=0.03 (referred to the fragment's center of mass). As we will see later on
in Sec. \ref{FB-systematcis}, oblate deformed fragments also
appear as a result of fissioning other  Uranium isotopes. Similar results have been obtained
in a recent study \cite{Robledo-Giulliani} based on the BCPM-EDF \cite{BCPM}. They deserve
further analysis, as it is usually assumed \cite{Moller-1,Moller-2} that fission fragments only exhibit prolate 
deformations. In our calculations, the deformed oblate fragment acquires this shape in order
to minimize a large Coulomb repulsion of 195.19 MeV. The 2F solution shown
is the one that minimizes the energy with the given quadrupole constraint. This
does not necessarily mean that this is the configuration obtained after scission.
In fact, it is observed experimentally that the mass number of the heavy fragment
is close to 140 instead of the $^{132}$Sn obtained as the minimum energy solution.
Successful theories of scission \cite{Wil.76} postulate that the breaking of 
the nucleus takes place when the neck between fragments reaches some 
critical value. If we consider the rupture point as the position where
the neck reaches its smallest width we obtain for the heavy fragment the
values $Z= 51.9$ and $N=84.5$  which are close to $Z=50$ and $N=82$ of
$^{132}$Sn but lead to a mass of 136.4 which is closer to the
experimental value. It has to be stressed that the values obtained should be taken 
as an approximation to the peaks of the  mass distribution of the fragments. Obviously, in 
order to reproduce the broad mass fragment distribution observed experimentally 
a dynamical theory considering the quantum mechanic evolution like, for instance,
the one of Ref.  \cite{Goutte-dynamical-distribution} is required.

% ----------------------------------------------------------------------
%
%  S U B S E C T I O N
%
%      H e a v y   n u c l e i   w i t h   k n o w n   
%
%      e x p e r i m e n t a l   d a t a 
%
% ----------------------------------------------------------------------  

\subsection{Heavy nuclei with known experimental data} 
\label{NucleiWithData}

In this section, the results obtained with the Gogny-D1M  EDF for the set of nuclei 
$^{232-238}$U, $^{238-244}$Pu, $^{240-248}$Cm, $^{250,252}$Cf, $^{250-256}$Fm, $^{252-256}$No, $^{256-260}$Rf, 
$^{258-262}$Sg, $^{264}$Hs and $^{286}$Fl are discussed. The selected nuclei
correspond to those where experimental data  are 
available \cite{Refs-barriers-other-nuclei-1,Refs-barriers-other-nuclei-2,Refs-barriers-other-nuclei-3-tsf}. 
Previous theoretical results, based on the parametrization D1S, have already been presented in
Refs. \cite{Warda-Egido-Robledo-Pomorski-2002,Delaroche-2006,Warda-Egido-2012}.
 
In Table I, we compare the predicted heights for the inner 
$B_{I}^\mathrm{th}$  and outer $B_{II}^\mathrm{th}$ barriers as well 
as the excitation energies $E_{II}^\mathrm{th}$ of the  fission 
isomers with the  experimental ones $B_{I}^\mathrm{exp}$, 
$B_{II}^\mathrm{exp}$ and $E_{II}^\mathrm{exp}$  
\cite{Refs-barriers-other-nuclei-1,Refs-barriers-other-nuclei-2}. The 
theoretical values have been obtained from the corresponding 
energies $E_{HFB}+ \Delta E_{ROT}$ by looking at the energy 
differences between the ground state energy and the energies of the 
corresponding maxima and minima. The axial $B_{I}^\mathrm{th}$ 
values are larger than the experimental ones 
\cite{Refs-barriers-other-nuclei-1} reaching a maximal  deviation 
$B_{I}^\mathrm{th}-B_{I}^\mathrm{exp}$ = 4.88 MeV in $^{248}$Cm. In 
order to explore the impact of the $\gamma$ degree of freedom, for 
all the nuclei reported in Table I, we have performed triaxial 
calculations for configurations with 10 b $\le$ $Q_{20}$ $\le$ 40 b. 
The parameter $\gamma$ takes on the values 0$^{o}$ $\le$ $\gamma$ 
$\le$ 12$^{o}$ in this range of quadrupole deformations. As can be 
seen from panels (a) to (n) of Fig. \ref{Heavy-Nuclei-gamma} and the 
numerical values given in parenthesis in the table, the triaxial 
inner barriers are systematically lower than the axial ones by up to 
3.18 MeV in $^{248}$Cm.  They also  display the position of the top 
of the barrier  shifted to higher quadrupole deformations. The 
triaxial heights still overestimate the experimental ones. However, 
having in mind that the Gogny-D1M EDF has not been fine tuned to 
fission data and the large uncertainties in the extraction of the 
experimental inner  barrier heights (an 1 MeV error bar is usually
presumed), it is more important that the 
global trend observed in the experiment and other theoretical models 
(see, for example, Refs. 
\cite{Robledo-Giulliani,UNEDF1,Abu-2012-bheights} and references therein) 
is reasonably well reproduced. The D1M values for  
$B_{I}^\mathrm{th}$ are consistent with the ones obtained in the 
framework of Gogny-D1S calculations \cite{Delaroche-2006}.

%%%%%%%%%%%%%%%%%%%%%%%%%%%%%%%%%%%%%%%%%%%%%%%%%%%%%%%%%%%%%%%%%%%%%%%%%%%%%%%%%%%%%%%%%%%%%%%%%
%
%   FIGURE OF THE PAPER  
%
%%%%%%%%%%%%%%%%%%%%%%%%%%%%%%%%%%%%%%%%%%%%%%%%%%%%%%%%%%%%%%%%%%%%%%%%%%%%%%%%%%%%%%%%%%%%%%%%%
\begin{figure}
\includegraphics[width=0.5\textwidth]{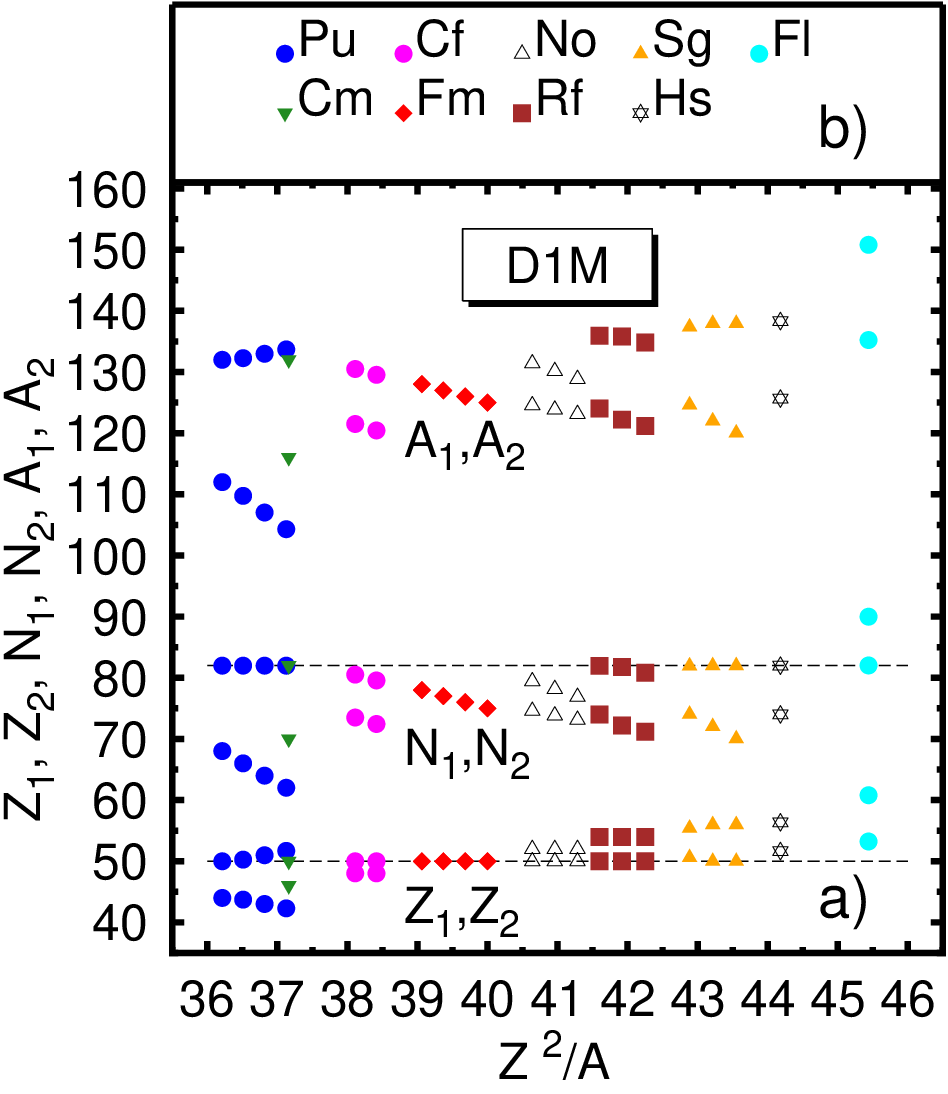}
\caption{ 
(Color online) 
In panel a), the proton ($Z_{1},Z_{2}$), neutron ($N_{1},N_{2}$) and mass ($A_{1},A_{2}$)
numbers of the two  fragments resulting from the fission of 
$^{238-244}$Pu, $^{248}$Cm, $^{250,252}$Cf, $^{250-256}$Fm, $^{252-256}$No, $^{256-260}$Rf, $^{258-262}$Sg, $^{264}$Hs 
and $^{286}$Fl  [see, panel b)] are shown as functions of 
the fissibility parameter $Z^{2}/A$ in the  parent nucleus. 
Results have been obtained with the Gogny-D1M EDF. The magic 
proton $Z=50$ and  neutron $N=82$ numbers
are highlighted with dashed horizontal lines to guide the eye.
}
\label{fragments-other-nuclei} 
\end{figure}
%%%%%%%%%%%%%%%%%%%%%%%%%%%%%%%%%%%%%%%%%%%%%%%%%%%%%%%%%%%%%%%%%%%%%%%%%%%%%%%%%%%%%%%%%%%%%%%%%

In the case of the 
outer barriers, the inclusion of reflection-asymmetric shapes  
leads to a reduction of a few MeV. However, we still observe deviations 
of up to $B_{II}^\mathrm{th}-B_{II}^\mathrm{exp}$= 4.75 MeV with respect to
the experimental data. As no significant effects 
are expected  from triaxiality ours, as well as 
previous Gogny-D1S results \cite{Delaroche-2006}, seem to indicate that 
other effects not related to the mass moments may be required
to further decrease the predicted  $B_{II}^\mathrm{th}$ values. Whether it
is the pairing degree of freedom or effects associated to symmetry 
restoration or the collective dynamics is something that remains to be
explored. However, we have to keep in mind the model dependent character
of the experimental data for outer barriers heights that makes those
quantities less reliable than the corresponding fission half-lives
for a comparison with theoretical values. In the case of the fission 
isomer excitation energy,  the largest difference observed 
$E_{II}^\mathrm{th}-E_{II}^\mathrm{exp}$=1.37 MeV occurs for $^{242}$Pu.

In Fig. \ref{tsf-other-nuclei} we compare  the Gogny-D1M  $t_\mathrm{SF}$ values, obtained
for the nuclei
 $^{232-238}$U, $^{240}$Pu, $^{248}$Cm, $^{250}$Cf, $^{250-256}$Fm, $^{252-256}$No, $^{256-260}$Rf, 
 $^{258-262}$Sg, $^{264}$Hs and $^{286}$Fl within the GCM and ATDHFB schemes, with the  
 experimental data \cite{Refs-barriers-other-nuclei-3-tsf}. Results are
  shown for $E_{0}$=0.5, 1.0, 1.5 and 
 2.0 MeV, respectively (see, Sec. \ref{Theory-used}). 
 The effect of triaxiality has not been taken into account
 in the calculations. The experimental fission half-lives
 expand a range of 27 orders of magnitude. The theoretical predictions 
  display a  larger variability 
 depending on whether the GCM or ATDHFB scheme is used as well as on the 
 $E_{0}$ parameter. For example, differences of up to 12, 9, 7 and 5 orders of 
 magnitude occur in $^{232-238}$U, $^{238-244}$Pu, $^{248}$Cm 
 and $^{250}$Cf, for $E_{0}$=0.5 MeV.
 Such differences 
  become smaller for the heavier 
  Fm, No, Rf, Sg, Hs and Fl nuclei. On the other hand, increasing  $E_{0}$  
  leads to smaller $t_\mathrm{SF}$ in either of the two schemes.
  This reduction 
  is particularly pronounced in the case of nuclei with higher and 
  wider fission barriers. It is  satisfying to observe
  that both the GCM and ATDHFB Gogny-D1M schemes capture 
  the large reduction of $t_\mathrm{SF}$ observed experimentally when
  going from   $^{232}$U to $^{286}$Fl.
 
The comparison along isotopic chains reveals that the trend with 
neutron number is also reasonably well described. For the nuclei 
depicted in  Fig. \ref{tsf-other-nuclei}, both our Gogny-D1M and 
previous  \cite{Warda-Egido-Robledo-Pomorski-2002,Warda-Egido-2012} 
Gogny-D1S calculations  exhibit a similar trend as a function of the 
fissibility parameter $Z^{2}/A$. However, larger $E_{0}$ values are 
required  in our case to improve the comparison with the 
experimental data. This is not surprising, as in most cases the 
Gogny-D1M 1F curves display a gentler decline for the largest 
deformations.

%-----------------------------------------------------------------------
%   F I G U R E :     U r a n i u m      D 1 S   
%-----------------------------------------------------------------------
\begin{figure*}
\includegraphics[width=1.0\textwidth]{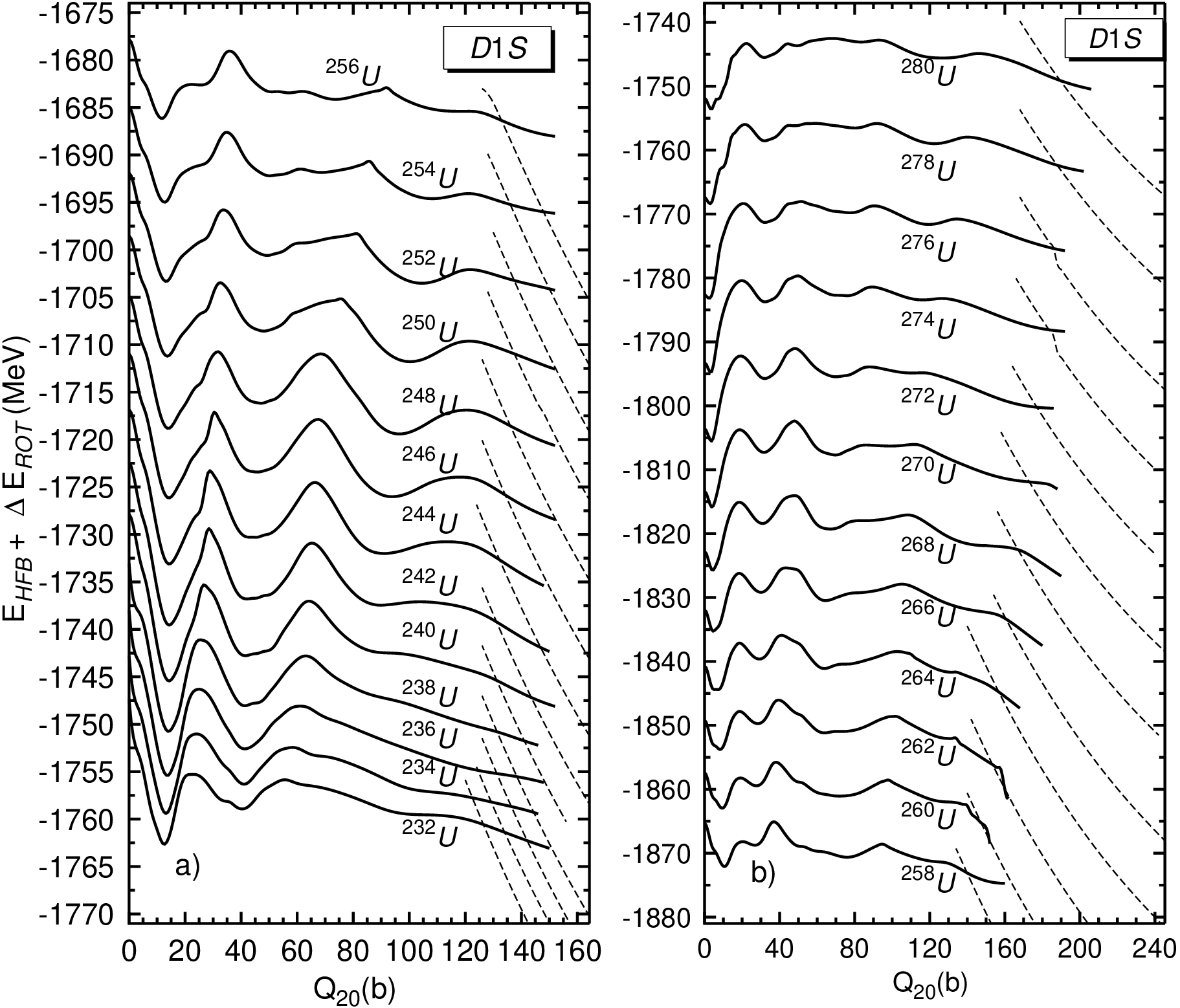} 
\caption{  
The HFB plus the zero point rotational energies obtained with the 
Gogny-D1S EDF are plotted in panel a) for the nuclei $^{232-256}$U 
and in panel b) for the nuclei $^{258-280}$U as functions of the quadrupole 
moment $Q_{20}$. Both the one (1F) and two-fragment (2F) solutions 
are shown  in the plot with continuous and dashed lines, respectively. 
Starting from $^{232}$U ($^{258}$U) in panel a) [in panel b)] all
the curves have been successively shifted by 15 MeV in order to accommodate
them in a single plot.  Note, that the energy scales span different 
ranges in each panel. For more details, see the main text. 
}
\label{FissionBarriersD1S} 
\end{figure*}
% -----------------------------------------------------------------------

The proton ($Z_{1},Z_{2}$), neutron ($N_{1},N_{2}$) and mass ($A_{1},A_{2}$)
numbers of the 2F solution leading to the minimum energy for
a given quadrupole moment and corresponding to the nuclei 
$^{238-244}$Pu, $^{248}$Cm, $^{250,252}$Cf, $^{250-256}$Fm, $^{252-256}$No, 
$^{256-260}$Rf, $^{258-262}$Sg, $^{264}$Hs  and $^{286}$Fl are shown in 
Fig. \ref{fragments-other-nuclei}, as functions of 
the fissibility parameter $Z^{2}/A$ of the  parent nucleus. 
Fragment properties have been obtained from the 2F solutions and for the
largest $Q_{20}$ values available as to guarantee that those properties are nearly 
independent of the quadrupole moment (which is equivalent to fragment's separation
for 2F solutions) considered.
In our calculations, the  proton and neutron numbers in the fragments are mostly 
dominated by the $Z=50$ and  $N=82$ magic numbers. Experimentally
\cite{Pu-mass-fragments-exp-1,Pu-mass-fragments-exp-2},  the average 
masses ${\overline{A}}_{H}$ of the heavy fission fragments in 
$^{238-244}$Pu, $^{248}$Cm, $^{250,252}$Cf and  $^{254,256}$Fm are nearly
constant with a value around  ${\overline{A}}_{H}$=140  and deviations of 1 or 2 mass 
units. As mentioned before, the 2F solution discussed here is 
determined by the minimum energy requirement and according to several 
models of scission this is not necessarily the configuration 
obtained after the break up of the parent nucleus.
In the previous section, we briefly mentioned that if the breakup point is
taken as the point where the well developed neck attains its minimum width then
the mass distribution becomes closer to the experimental values. However, a more
microscopic model including quantum-mechanical effects like the one 
of Ref. \cite{Goutte-dynamical-distribution} should be used for a 
sounder theoretical description. As this kind of dynamical
model is very involved computationally we will not dwell on this and we just
keep in mind that the mass distribution of the two fragments leading to the
minimum energy at the HFB level underestimates the mass of the heavier fragment by
a few units. In addition to this general consideration we can encounter locally
examples where our model is not able to reproduce the delicate balance
between macroscopic and shell effects that lead, for example, to 
mass asymmetric splittings in the heavy Fm isotopes. As an example, 
let us mention that a symmetric splitting  is obtained in  
$^{256}$Fm in disagreement with the rather large mass asymmetry  
${\overline{A}}_{L}/{\overline{A}}_{H}$=112/141 observed experimentally. 
Similar results have  been obtained in previous calculations with 
the Gogny-D1S EDF \cite{Warda-Egido-Robledo-Pomorski-2002}. 
On the other hand, the ratio ${\overline{A}}_{L}/{\overline{A}}_{H}$=124/136 
predicted for  $^{260}$Rf coincides with the one  reported in 
Ref. \cite{Warda-Egido-Robledo-Pomorski-2002}.

To summarize the conclusions of this section, it has been shown that 
in spite of large theoretical uncertainties in the choice of the 
models to describe the relevant quantities, the Gogny-D1M 
\cite{gogny-d1m} HFB framework provides a reasonable description of the 
tendencies with mass number of the physical observables. This 
validates the use of this parametrization to study the systematics 
of fission paths and other relevant quantities in the isotopes 
$^{232-280}$U that is presented in the next section. Results 
obtained with the D1S  and D1N parameter sets will also be discussed to quantify the 
typical uncertainties associated to the employed Gogny-EDF.

%-----------------------------------------------------------------------
%   F I G U R E :     U r a n i u m      D 1 N   
%-----------------------------------------------------------------------
\begin{figure*}
\includegraphics[width=1.0\textwidth]{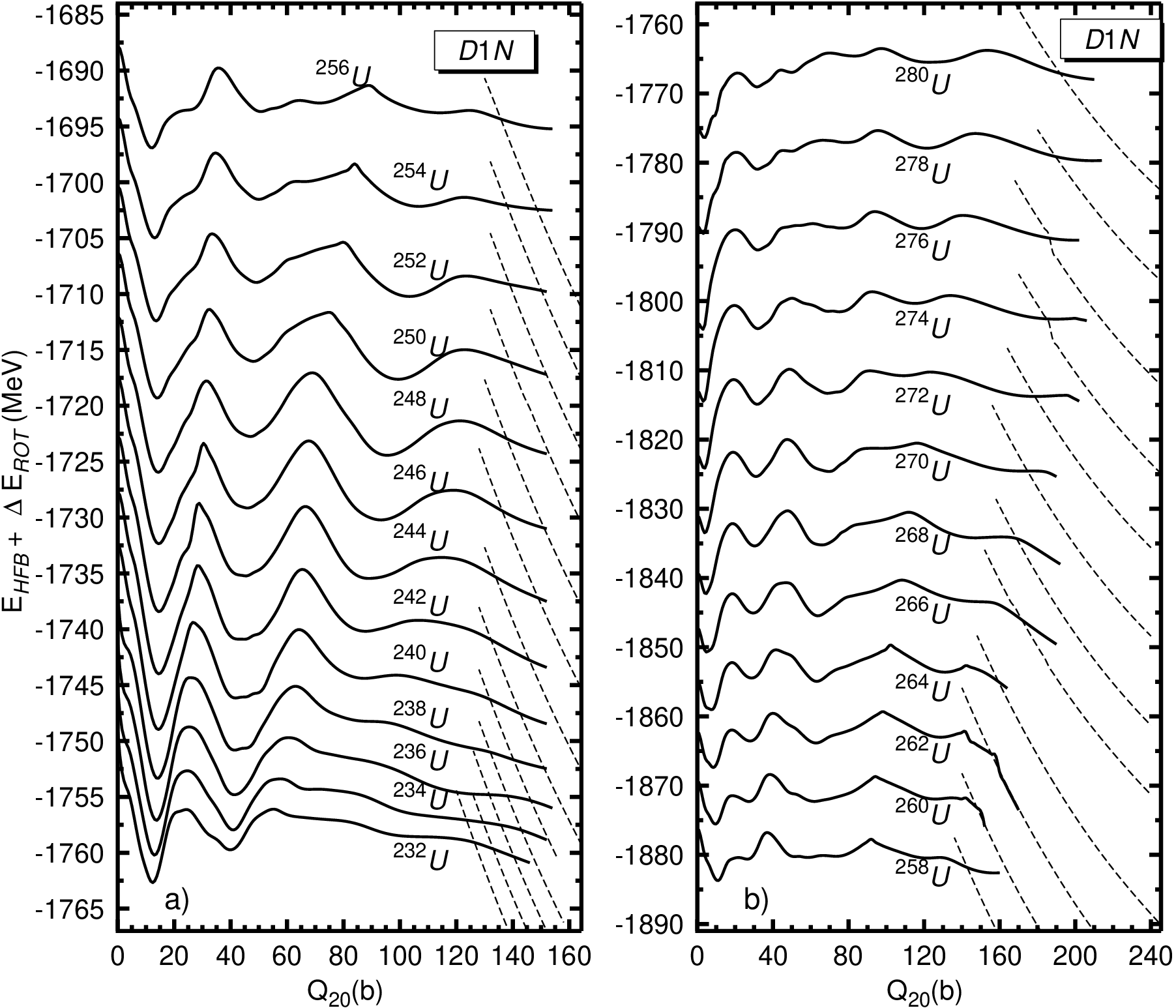} 
\caption{ 
The same as Fig. \ref{FissionBarriersD1S} but for the Gogny-D1N EDF.
}
\label{FissionBarriersD1N} 
\end{figure*}
%-----------------------------------------------------------------------

% -----------------------------------------------------------------------
%
%   N e u t r o n   r i c h    U r a n i u m    i s o t o p e s
%
%-----------------------------------------------------------------------

\subsection{Systematics of fission paths, spontaneous fission half-lives and 
fragment mass in Uranium isotopes}
\label{FB-systematcis}

\begin{figure*}
\includegraphics[width=1.0\textwidth]{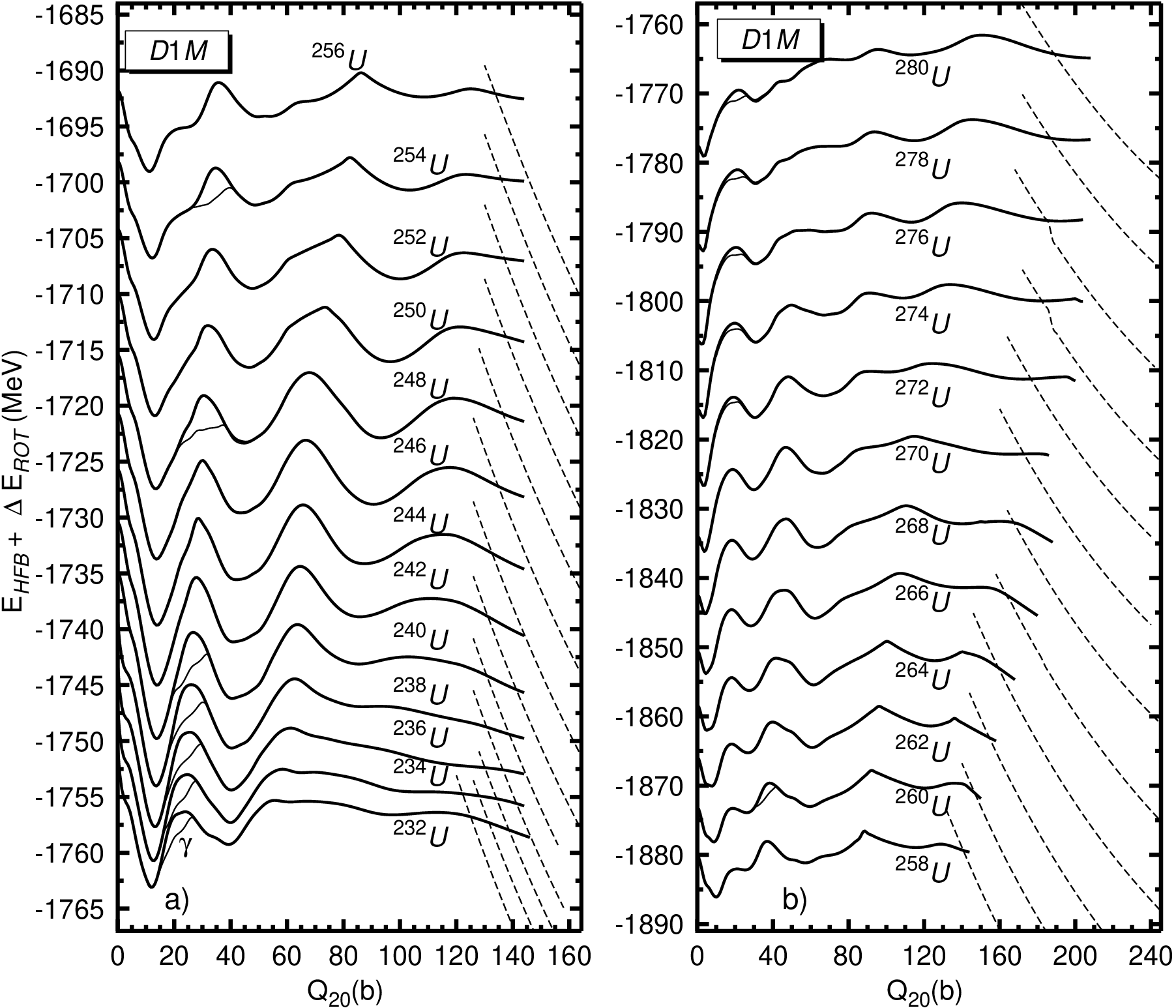} 
\caption{ 
The same as Fig. \ref{FissionBarriersD1S} but for the 
Gogny-D1M EDF. The energies obtained in the framework of triaxial 
calculations for the nuclei $^{232-240,248,254,260,272-280}$U 
(labeled as $\gamma$) are also included in the plot.
}
\label{FissionBarriersD1M} 
\end{figure*}

In Figs. \ref{FissionBarriersD1S}, \ref{FissionBarriersD1N} and 
\ref{FissionBarriersD1M} we have plotted the energies $E_{HFB} + \Delta E_{ROT}$, obtained 
with the D1S, D1N and D1M Gogny-EDFs, for the nuclei $^{232-256}$U  [panel (a)] and 
$^{258-280}$U [panel (b)]. Both the  1F (full lines)  and 2F (dashed lines) curves  
are shown in the plots. Starting from $^{232}$U ($^{258}$U) in panel (a) [in panel (b)] all
the curves have been successively shifted by 15 MeV in order to 
accommodate them in a single plot. Before commenting on more quantitative
aspects of the results it is worth to notice that, regardless of the 
functional employed, the shapes of the 1F and 2F curves in $^{232-280}$U
look rather similar pointing to equivalent liquid drop and shell effect 
properties of the three EDFs considered.

As the neutron number increases, we observe a gradual decrease
in the deformations corresponding to the absolute minima of 
the 1F curves in Figs. \ref{FissionBarriersD1S}, \ref{FissionBarriersD1N} and 
\ref{FissionBarriersD1M} reaching the value $Q_{20} \approx $ 4 b
in the heavier isotopes. The non-zero $Q_{20}$ value of the ground
state minimum  in $^{270-280}$U is a direct consequence of the 
rotational energy correction that shifts spherical HFB minima to non-zero
quadrupole moments, as it has  already been documented 
in several regions of the nuclear chart \cite{RRG23S,ER-Lectures,NPA-2002,Rayner-PRC2004}.
If just the HFB energy is considered these nuclei have a spherical intrinsic
state consequence of neutron numbers close to $N=184$ that is predicted 
to be a magic number in our calculations (see below).

An increase in the height of the inner fission barriers and the 
widening of the 1F curves is noticed in all the considered 
Gogny-EDFs as the two-neutron dripline is approached. As a 
consequence, an increase in the spontaneous fission half-lives for 
the heavier Uranium isotopes is expected. It is also worth 
mentioning, the existence of second fission isomers in several of 
the considered nuclei. For example, in $^{240 - 252}$U they exhibit 
quadrupole deformations $Q_{20} \approx$ 86-96 b. The second fission 
isomers are also visible in the 1F curves of heavier isotopes though 
in some cases the situation is not  as well defined due to the 
presence of several shallow minima. Similar results have been 
recently obtained with the BCPM-EDF \cite{Robledo-Giulliani}.

In order to explore the role of the $\gamma$ degree of freedom, we 
have performed Gogny-D1M triaxial calculations, for the isotopes 
$^{232-240,248,254,260,272-280}$U. The corresponding results for 
$^{240}$U and $^{234-238}$U have already been shown in Figs. 
\ref{FissionBarrierD1M240Uplusgamma} and \ref{Heavy-Nuclei-gamma}, 
respectively, but they are included again in Fig. 
\ref{FissionBarriersD1M} for the sake of completeness. For the heavier  
nuclei $^{248,254,260,272-280}$U, we have performed triaxial 
calculations for 4 b $\le$ $Q_{20}$ $\le$ 50 b, with $\gamma$= 20
$^{o}$ being the largest value considered. The corresponding 
energies are shown in Fig. \ref{FissionBarriersD1M} and thin lines 
visible in the neighborhood of the first fission barrier. 

%-----------------------------------------------------------------------
%   F I G U R E :     U r a n i u m      D 1 M
%-----------------------------------------------------------------------
%-----------------------------------------------------------------------

\begin{figure}
\includegraphics[width=1.0\columnwidth]{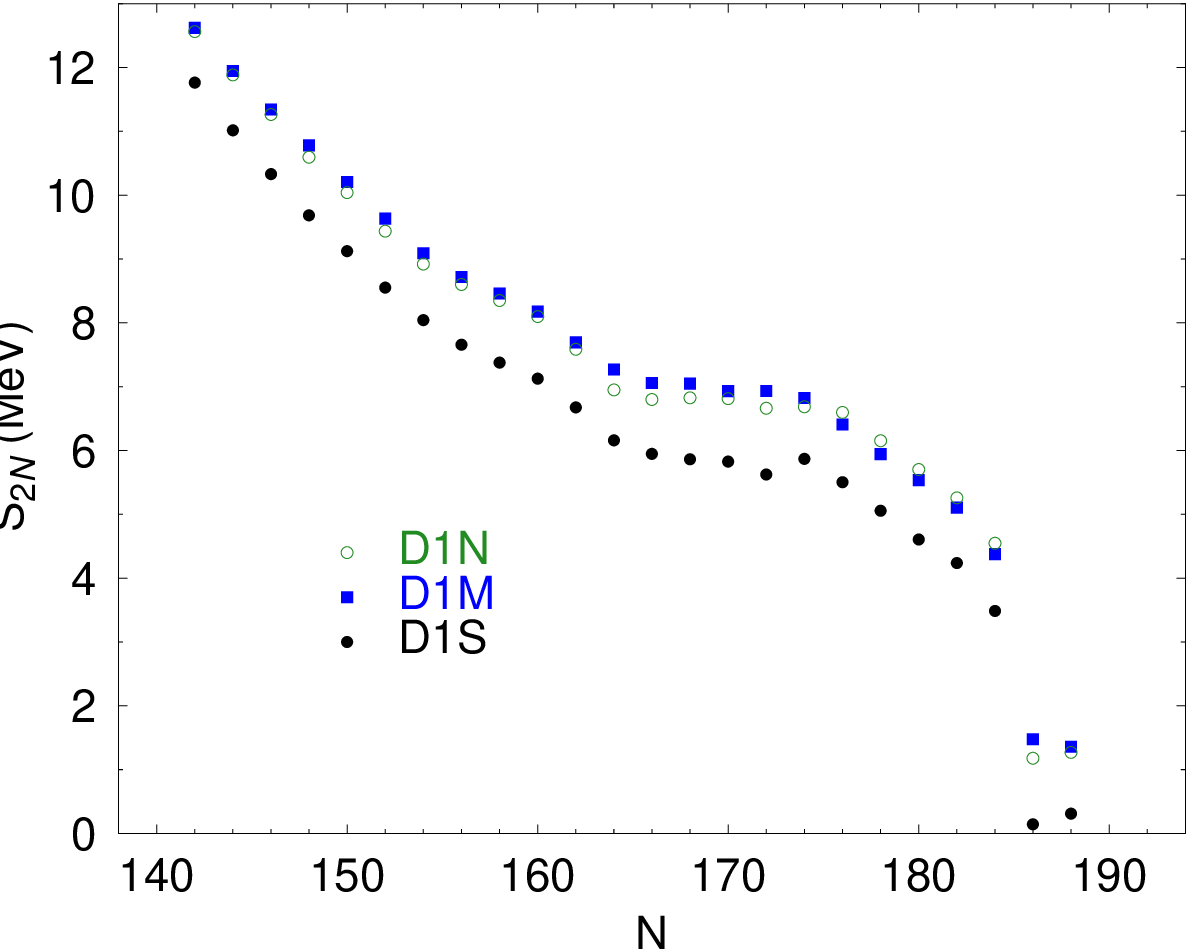} 
\caption{ 
(Color online) Two neutron separation energies $S_{2N}$ as a 
function of neutron number
}
\label{S2N} 
\end{figure}

In order to better understand the trends in binding energies for the 
Uranium isotopes the two neutron separation energies ($S_{2N}$) are 
plotted in Fig. \ref{S2N} for the three sets of calculations.  We 
observe that whereas the D1N and D1M $S_{2N}$ are rather similar, 
the D1S values are typically 1 MeV lower than the previous ones. 
These low values for D1S were reported in previous large-scale 
calculations \cite{Hilaire-systematics-D1S} and show up as a 
systematic drift in the differences between the experimental and 
theoretical binding energies in heavy nuclei. In fact, the effort to 
correct this drift in the quest for an accurate mass table based on 
the  Gogny-EDF led to the proposal of both D1N \cite{gogny-d1n} and 
D1M \cite{gogny-d1m}. Previous  studies 
\cite{PTpaper-Rayner,gogny-d1n,gogny-d1m,Rayner-Sara,Rayner-Robledo-JPG-2009,Rayner-PRC-2010,Rayner-PRC-2011} 
suggest that, while improving the 
description of nuclear masses, both the D1N and D1M sets still have 
the same essential predictive power to describe low-energy nuclear 
structure properties as the Gogny-D1S EDF. Nevertheless, more 
calculations are still required to substantiate this conclusion. The 
main features observed in the $S_{2N}$ are the plateau between $N=166$
and $N=174$ and the sudden drop at $N=186$ that signals the magic 
number character of $N=184$. 

%-----------------------------------------------------------------------
% F I G U R E :    T w o   n e u t r o n   s e p a r a t i o n 
%                  e n e r g i e s 
%-----------------------------------------------------------------------
%-----------------------------------------------------------------------

\begin{figure}
\includegraphics[width=1.0\columnwidth]{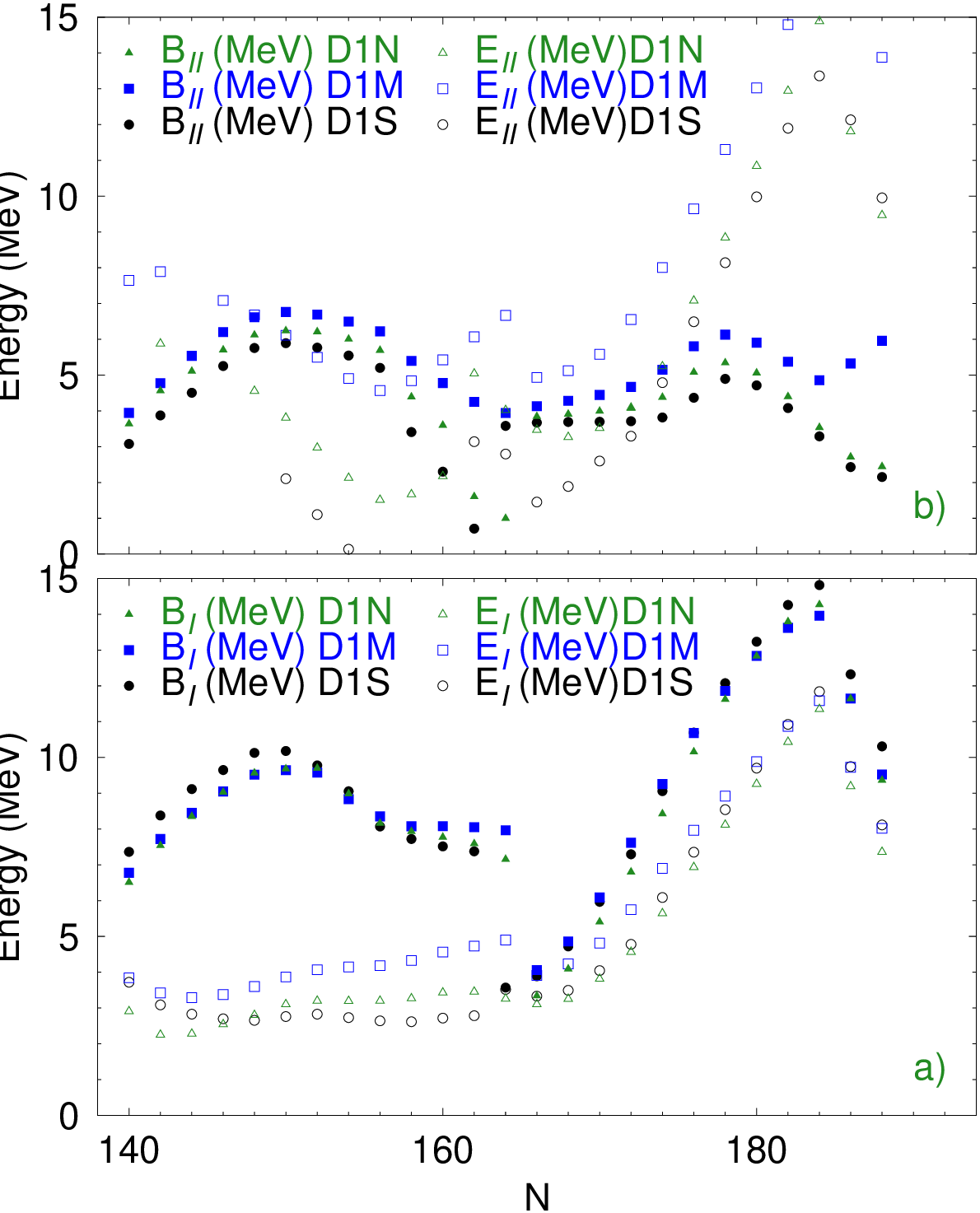} 
\caption{
(Color online) Excitation energies $E_{I}$ ($E_{II}$) and
fission barrier heights $B_{I}$ ($B_{II}$) for the first (second) well 
are shown in panels a) [b)].
}
\label{EBI} 
\end{figure}

In Fig. \ref{EBI} the excitation energies of the
first $E_{I}$  fission isomers are plotted along with the first barrier height 
$B_{I}$ in panel a). In panel b) the same quantities but referred to the
second isomeric well  are shown. The results for the three EDFs show a similar
behavior with neutron number for the first barrier height $B_{I}$. A sudden
drop is observed at $N=164$ that is correlated with the plateau observed
in the $S_{2N}$ plot. The excitation energies of the first fission isomer
remain more or less constant with a value around 3 MeV up to $N=164$ where
they show a change of tendency and start to increase linearly with neutron
number. At $N=184$ there is a sudden drop accompanied by a drop in $B_{I}$
characteristic of the filling of a new major shell. 
As previously mentioned, for D1M triaxial calculations in the neighborhood of
the inner fission barriers have been carried out. Triaxiality reduces
the $B_{I}$ by 0.55, 0.59, 1.33, 1.60 and 1.96 MeV in the case of 
$^{232-240}$U, respectively. In spite of this reduction the theoretical 
predictions still overestimate the experimental data
\cite{Refs-barriers-other-nuclei-1,Refs-barriers-other-nuclei-2}
for $^{234-238}$U (see, also Table I). On the 
other hand, the axial barrier heights 8.30, 8.02, 8.44, 12.71, 13.32, 13.94, 11.42 and 9.54 MeV 
 in the nuclei 
$^{248,254,260,272-280}$U are reduced by
 2.70, 1.75, 0.56, 1.33, 0.86, 1.04, 1.01 and 0.88 MeV, respectively.

For the second isomeric well the behavior is more erratic and we can
even observe the lack of second isomeric well in some nuclei. It is also
worth noticing the similar predictions for $B_{II}$ from the three EDFs 
and the large dispersion in the predicted $E_{II}$ values.

%-----------------------------------------------------------------------
% F I G U R E :    B a r r i e r   h e i g t h s   a n d   f i s s i o n
%                  i s o m e r    e n e r g i e s 
%-----------------------------------------------------------------------
%-----------------------------------------------------------------------

\begin{figure*}
\includegraphics[width=1.0\textwidth]{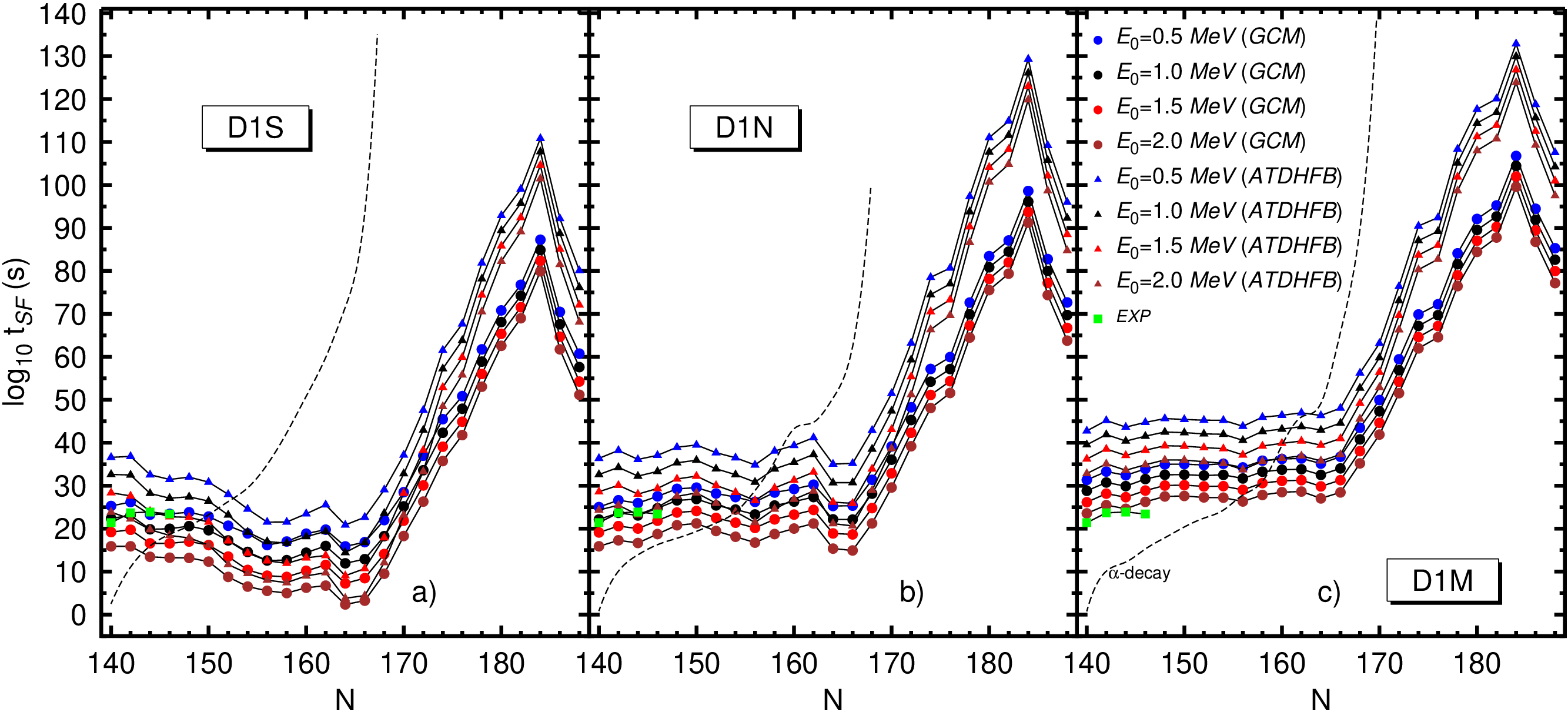} 
 \caption{ 
(Color online) The spontaneous fission half-lives $t_\mathrm{SF}$, 
predicted within the GCM and ATDHFB schemes, for the isotopes 
$^{232-280}$U  are depicted as functions of the neutron number. 
Results have been obtained with the Gogny-D1S [panel a)], Gogny-D1N 
[panel b)] and Gogny-D1M [panel c)] EDFs. For each parametrization, 
calculations have been carried out  with $E_{0}$=0.5, 1.0, 1.5 and 
2.0 MeV, respectively. The experimental $t_\mathrm{SF}$ values for 
the nuclei $^{232-238}$U are included in the plot. In addition,  
$\alpha$-decay half-lives are plotted with  short dashed lines. For 
more details, see the main text.
 }
\label{tsf-D1S-D1N-D1M} 
\end{figure*}

With all the previous ingredients at hand, we have computed the 
spontaneous fission half-lives Eq. (\ref{TSF}) for  the considered 
Uranium isotopes. The effect of triaxiality, is not taken into 
account in the calculations. The $t_\mathrm{SF}$ values, predicted 
within the GCM and ATDHFB schemes, are plotted in Fig. 
\ref{tsf-D1S-D1N-D1M} as a function of the neutron number. Results have 
been obtained with the Gogny-D1S [panel (a)], Gogny-D1N [panel (b)] 
and Gogny-D1M [panel (c)] EDFs. For each parametrization, we have 
carried out calculations with $E_{0}$=0.5, 1.0, 1.5 and 2.0 MeV, 
respectively. The experimental $t_\mathrm{SF}$ data for  $^{232-238}$
U are  included in the plot. 

%-----------------------------------------------------------------------
% F I G U R E :    F i s s i o n   h a l f   l i v e s    U r a n i u m 
%                  i s o t o p e s.                    
%-----------------------------------------------------------------------
%-----------------------------------------------------------------------

The $t_\mathrm{SF}$ values  predicted within the ATDHFB 
approximation are always larger than the GCM ones for a given $E_{0}$.
For example, for $E_{0}$ =0.5 MeV, differences of up to 12  orders 
of magnitude are obtained for the lighter isotopes. Such differences 
increase with increasing neutron number reaching 23, 31 and 26 
orders of magnitude in the nucleus $^{276}$U with the 
parametrizations D1S, D1N and D1M, respectively. Increasing $E_{0}$ 
leads always to a decrease in $t_\mathrm{SF}$. It is satisfying to 
observe that all the parametrizations lead to the same trend in 
$t_\mathrm{SF}$, even though D1M provides the largest absolute 
values in half-lives and, as already discussed in Sec. \ref{NucleiWithData}, 
larger $E_{0}$ values are required to improve the agreement with the 
available experimental data. This is a consequence of the shape of 
the 1F curves provided by the Gogny-D1M EDF for the considered 
Uranium isotopes that are wider than for the other EDFs. Regardless 
of the EDF employed, we observe a steady increase in the spontaneous 
fission half-lives for neutron numbers $N \ge 166$ reaching a 
maximum at $N=184$, which is predicted to be a magic neutron number in 
our calculations.

In Fig. \ref{tsf-D1S-D1N-D1M}, we have also plotted the $\alpha$-decay half-lives 
computed with a parametrization \cite{TDong2005} 
of the Viola-Seaborg formula \cite{Viola-Seaborg} Eq. 
(\ref{VSeaborg-new}). To this end, we have used the  binding energies 
obtained for the corresponding Uranium and Thorium isotopes [see, 
Eq. (\ref{QALPHA})]. Here, we stress that, at variance with the  
Gogny-D1S \cite{gogny-d1s}, both the D1N \cite{gogny-d1n} and D1M 
\cite{gogny-d1m} parametrizations have been tailored to give a 
better description of the nuclear masses and therefore their $\alpha$
decay half-lives are expected to be much more realistic than the 
D1S ones. In all cases, a steady increase is observed in  
$t_{\alpha}$  as a function of the neutron number. Though the 
precise value depends on the selected EDF (i.e., $N=144$ for D1S, 
$N=150$ for D1N and $N=156$ for D1M), it is clearly seen that for 
increasing neutron number fission turns out to be faster than 
$\alpha$-decay. 

Our predictions compare well with the semiclassical results of Ref \cite{mam01}
using the Extended Thomas-Fermi method for a Skyrme interaction. In that
calculation very high barriers are predicted for $N=184$ in the uranium
isotopic chain contrary to some liquid drop models. The barrier heights
in those neutron rich nuclei are correlated to the surface symmetry 
energy coefficient $a_{ss}$, an effect that deserves further study for
the Gogny class of energy functionals.

\begin{figure*}
\includegraphics[width=1.0\textwidth]{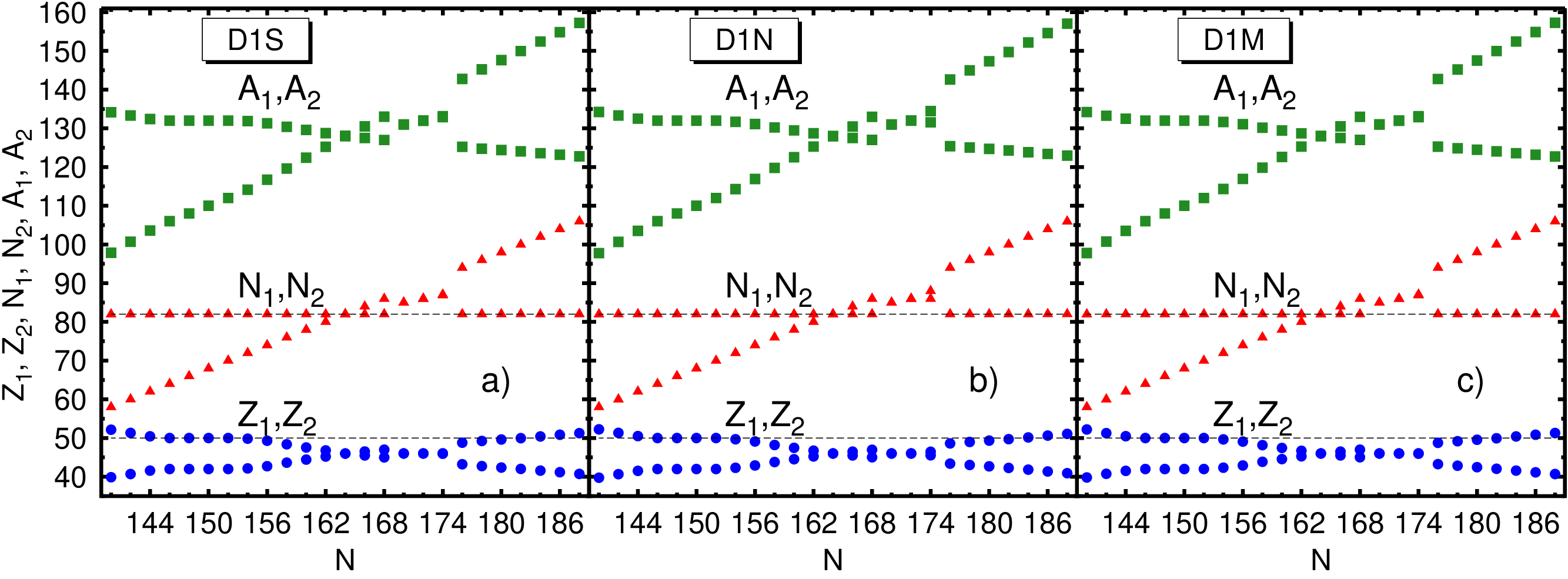} 
\caption{ 
(Color online) The proton ($Z_{1},Z_{2}$), neutron ($N_{1},N_{2}$) and 
mass ($A_{1},A_{2}$) numbers of the two  fragments resulting from the 
fission of the isotopes $^{232-280}$U are shown as functions of the 
neutron number in the  parent nucleus.
Results have been obtained with the Gogny-D1S [panel a)], 
Gogny-D1N [panel b)] and Gogny-D1M [panel c)]
EDFs. The magic proton $Z=50$ and  neutron $N=82$ numbers
are highlighted with dashed horizontal lines to guide the eye.  
}
\label{mass-D1S-D1N-D1M} 
\end{figure*}

The proton ($Z_{1},Z_{2}$), neutron ($N_{1},N_{2}$) and mass 
($A_{1},A_{2}$) numbers of the 2F solutions for 
$^{232-280}$U are shown in Fig. \ref{mass-D1S-D1N-D1M} as 
functions of the neutron number of the parent nucleus. Results have been 
obtained with the Gogny-D1S [panel (a)], Gogny-D1N [panel (b)] and 
Gogny-D1M [panel (c)] EDFs. Exception made of the nuclei $^{262-266}$
U, the neutron number in one of the fragments always corresponds to 
the magic number $N=82$ while for the other fragment it increases as a 
function of the neutron number in the parent nucleus. The proton 
number in one of the fragments is always close to the magic one 
$Z=50$ for  $^{232-248}$U and $^{268-280}$U which, in the case of 
the light isotopes, agrees well with the experiment \cite{Schmidt}. 
It also varies almost linearly with the neutron number in the parent 
nucleus except for  $^{238-244}$U (it stabilizes at $Z=50$) and  
$^{256}$U (symmetric splitting). Note that a symmetric splitting is 
also predicted  for the nuclei $^{262,264}$U with the three EDFs. On 
the other hand,  both the  D1S and D1M 
parametrizations provide for $^{266}$U a symmetric splitting while a small 
difference in the neutron and proton numbers of the two fragments 
is  obtained with the Gogny-D1N EDF. The relevance of magic numbers 
in the fragments mass distribution is not surprising as it has been 
obtained by using minimum of the energy criteria. Therefore, as 
discussed previously, they are not directly comparable to the real 
fission mass distribution. 

%-----------------------------------------------------------------------
%  F I G U R E :     Z,  N   a n d   A   f r a g m e n t   m a s s 
%                    d i s t r i b u t i o n 
%-----------------------------------------------------------------------
%-----------------------------------------------------------------------

\begin{figure*}
\includegraphics[width=0.3\textwidth]{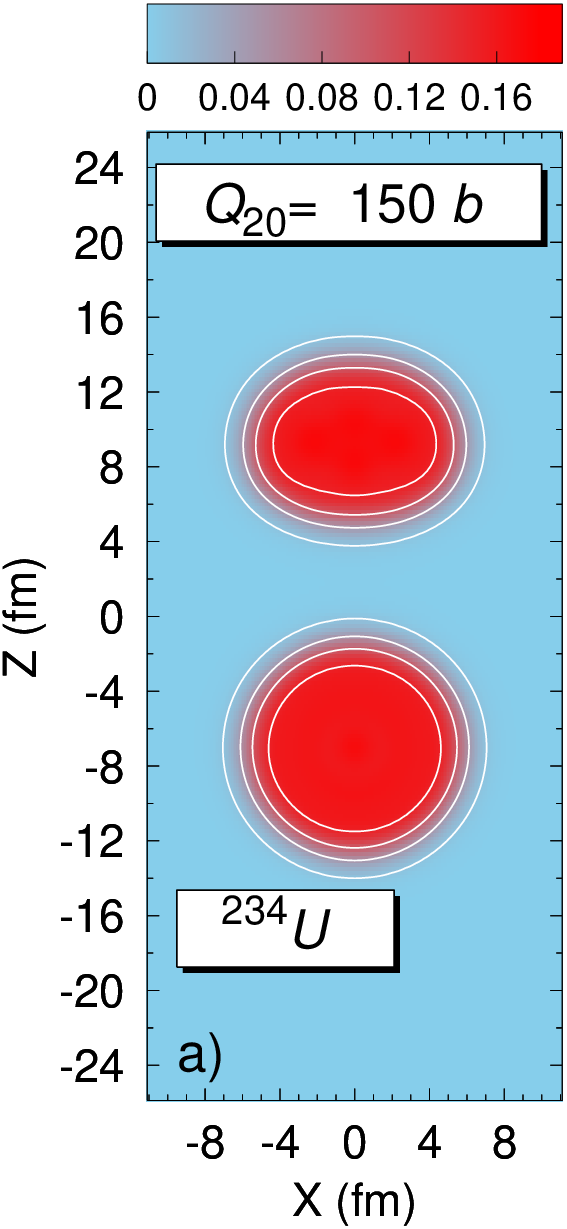}
\includegraphics[width=0.3\textwidth]{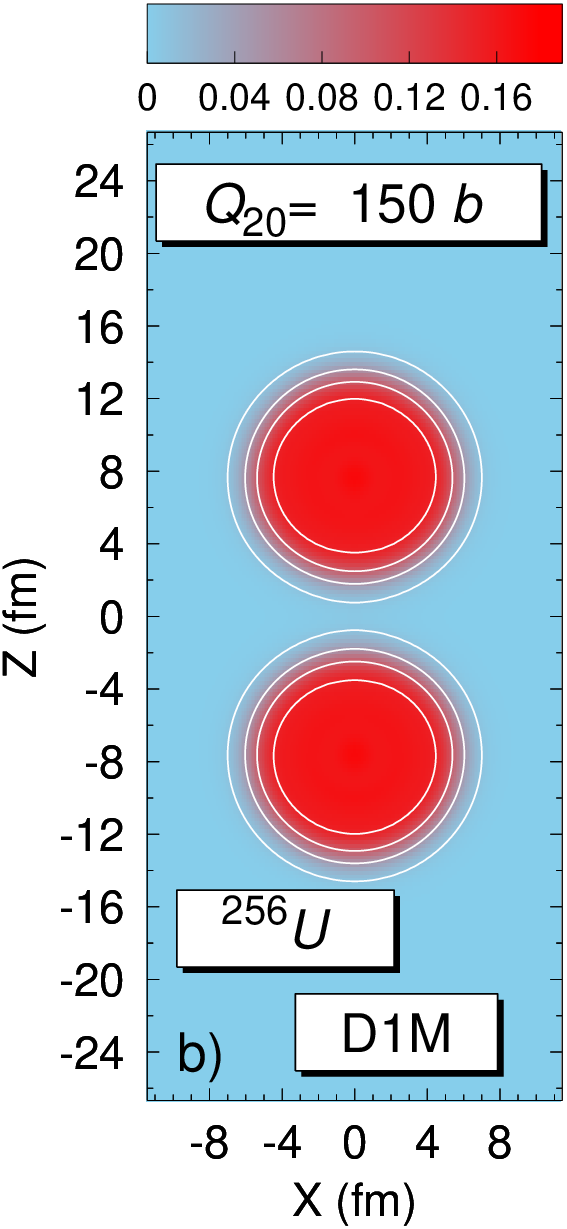}
\includegraphics[width=0.3\textwidth]{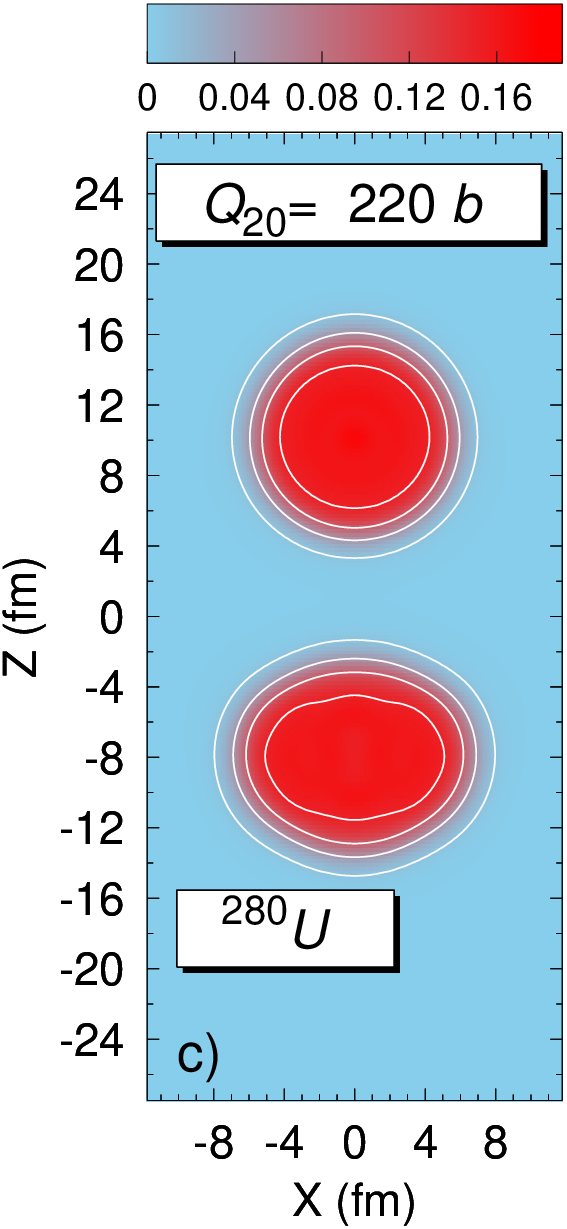}
\caption{ 
(Color online) Density contour plots for the nuclei
$^{234}$U [panel a)],  $^{256}$U [panel b)] an  $^{280}$U [panel c)].
The density profiles correspond to 2F solutions at the quadrupole 
deformations $Q_{20}$=150, 150 and 220 b, respectively. Results are 
shown for the parametrization D1M of the Gogny-EDF. The density is 
in units of fm$^{3}$ and contour lines correspond to densities
0.01, 0.05, 0.10 and 0.15 fm$^{-3}$. 
}
\label{den_cont_234-256-280U_D1M} 
\end{figure*}

We have also studied the evolution of the shapes in the fission 
fragments. A typical outcome of our calculations is shown in 
Fig. \ref{den_cont_234-256-280U_D1M}, where the density contour plots for
the nuclei $^{234}$U [panel (a)], $^{256}$U [panel (b)] and $^{280}$U [panel (c)]
are plotted at quadrupole deformations ($Q_{20}$=150, 150 and 220 b, 
respectively) corresponding to 2F solutions of the HFB equations 
(see, Fig. \ref{FissionBarriersD1M}).
Results are shown for the parametrization D1M but similar ones are 
obtained for the other Gogny-EDFs.  The lighter (heavier) fragment 
in $^{234}$U  ($^{280}$U) is predicted to be 
oblate and slightly octupole deformed with 
$\beta_{2}$=-0.22 and $\beta_{3}$=0.02. On the other hand, for the isotope
$^{256}$U we have obtained two identical fragments with 
$\beta_{2}$=-0.02 and $\beta_{3}$=0.01. As already mentioned in 
Sec. \ref{strategy-240U}, the appearance of  oblate  fragments in
our calculations deserves further attention as fission fragments
are usually assumed  \cite{Moller-1,Moller-2} to  be prolate deformed. 

The  results  discussed in this section, show that the same trends 
are obtained with the D1S, D1N and D1M parametrizations. This gives 
us confidence in the robustness of our predictions with respect to 
the version of the Gogny-EDF employed. In particular, from the 
previous results and the ones discussed in Sec. 
\ref{NucleiWithData}, we conclude that the Gogny-D1M EDF represents a 
reasonable starting point to describe fission properties in the  
isotopes $^{232-280}$U and other heavy nuclei.  With this in mind, 
we will proceed to explicitly discuss the impact of pairing 
correlations in the next Sec. \ref{change-pairing-strenght}.

%-----------------------------------------------------------------------
% S U B S E C T I O N :                V a r y i n g   p a i r i n g   
%                                      s t r e n g t h s
%-----------------------------------------------------------------------

\subsection{Varying pairing strengths in Uranium isotopes}
\label{change-pairing-strenght}

In this section, we discuss the impact of the strength of pairing 
correlations on the predicted spontaneous fission half-lives and 
other relevant fission properties in $^{232-280}$U. To this end, we 
have carried out  self-consistent calculations with a modified 
Gogny-D1M EDF in which, a multiplicative factor $\eta$ has been 
introduced in front of the HFB pairing field $\Delta_{kl}$ 
\cite{rs}. The corresponding pairing interaction energy reads 
\begin{eqnarray} \label{pairing-interaction-energies}
E_{pp}(\eta) = - \frac{\eta}{2} \mathrm{Tr} \left(\Delta \kappa^{*} \right)
\end{eqnarray}
For simplicity, we have considered the same $\eta$-factor for both protons 
and neutrons. In addition to the normal Gogny-D1M EDF (i.e., $\eta$=1), 
calculations have then been carried out with $\eta$= 1.05 and 1.10, 
respectively. Our main reason to consider different pairing 
strengths is that they are key ingredients in the 
computation of both the collective masses and the zero point 
energies. For example, it has already been shown 
\cite{proportional-1,proportional-2}
that the collective mass  is inversely proportional
to some power of the pairing gap, i.e., the stronger the pairing correlations are
the smaller the collective masses become. Similar $\eta$-factors have been 
recently used in Ref. \cite{Robledo-Giulliani} as well as 
to describe pairing and rotational properties of actinides and 
superheavy nuclei in the framework of the  RMF
approximation (see, for example, Ref. \cite{Afa-Abdu}
and references therein).

%%%%%%%%%%%%%%%%%%%%%%%%%%%%%%%%%%%%%%%%%%%%%%%%%%%%%%%%%%%%%%%%%%%%%%%%%%%%%%%%%%%%%%%%%%%%%%%%%
%
%   FIGURE OF THE PAPER  
%
%%%%%%%%%%%%%%%%%%%%%%%%%%%%%%%%%%%%%%%%%%%%%%%%%%%%%%%%%%%%%%%%%%%%%%%%%%%%%%%%%%%%%%%%%%%%%%%%%
%%%%%%%%%%%%%%%%%%%%%%%%%%%%%%%%%%%%%%%%%%%%%%%%%%%%%%%%%%%%%%%%%%%%%%%%%%%%%%%%%%%%%%%%%%%%%%%%%

\begin{figure}
\includegraphics[width=0.5\textwidth]{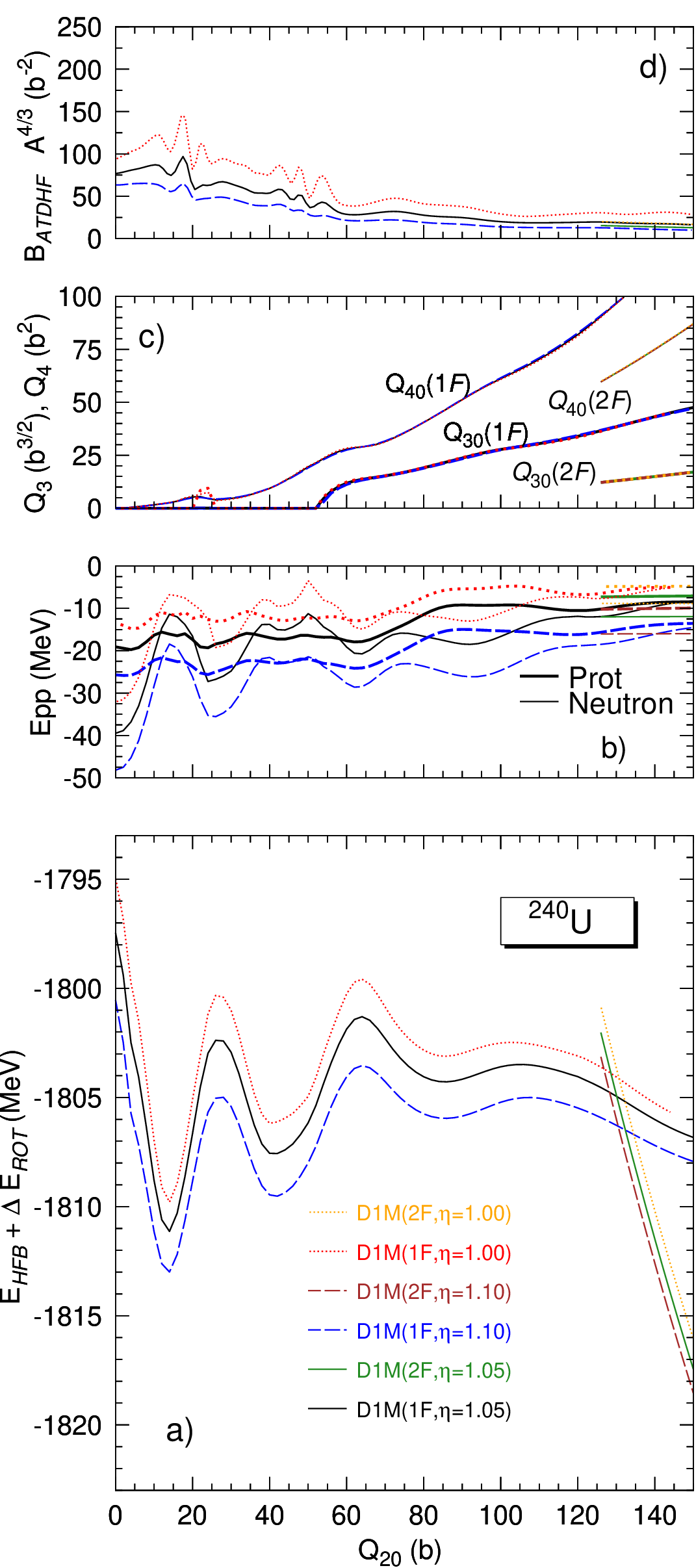}
\caption{ 
(Color online) The HFB plus the zero point rotational energies obtained 
with the normal ($\eta$=1.00) and modified ($\eta$=1.05 and 1.10)
Gogny-D1M EDFs are plotted in panel a) as 
functions of the quadrupole 
moment $Q_{20}$ for the nucleus $^{240}$U. For each 
$\eta$ value, both the one (1F) and two-fragment (2F) solutions 
are included  in the plot. The pairing interaction energies are depicted 
in panel b) for
protons (thick lines) and neutrons (thin lines). The octupole and hexadecapole moments 
corresponding to the 1F and 2F solutions are given in panel c). The collective masses
obtained within the ATDHFB approximation are plotted in panel d). For more
details, see the main text.
}
\label{FissionBarriers-eta-240U} 
\end{figure}

A typical outcome of our calculations is shown in Fig. 
\ref{FissionBarriers-eta-240U} (a) where, we compare the three fission 
profiles obtained for the nucleus $^{240}$U using the normal ($\eta$
= 1.00)  and modified  ($\eta$=1.05 and 1.10) Gogny-D1M EDFs. For 
each $\eta$ value, both the  1F and 2F solutions are included  in 
the plot. Exception made of the corresponding energy shifts, the 1F 
and 2F curves in $^{240}$U and all the other Uranium isotopes, 
exhibit rather similar energy shapes. The  ground state in $^{240}$U 
located around $Q_{20}$=14 b and its deformation decreases with 
increasing $\eta$. Increasing the pairing strength 
by 5  and 10 $\%$ we gain  1.11 and 2.29 MeV in binding energy, 
respectively. These quantities have to be compared to the HFB pairing
correlation energy of 1.92 MeV obtained by subtracting the HFB energy
to the Hartree-Fock one. We observe an increase of around $60\%$ in 
correlation energy for $\eta=1.05$ which is consistent 
with the exponential dependence of the correlation energy on the pairing
strength. In spite of the large impact on correlation energies
other quantities considered to fix the pairing strength like two neutron
separation energies do not change significantly when $\eta$ is increased
justifying the range of $\eta$ values considered. On the other hand, 
the heights of the inner barriers (8.76 MeV for $\eta$
=1.05 and 8.00 MeV for $\eta$=1.10) display a reduction of 720 KeV 
and 1.47 MeV when compared to the one obtained using the normal 
Gogny-D1M EDF. The excitation energy of the first 
fission isomer, located at $Q_{20}$=42 b, is  lowered by 50 KeV (
$\eta$= 1.05) and 140 KeV ($\eta$= 1.10). 

The proton (dashed lines) and neutron (full lines) pairing 
interaction energies are depicted in  Fig. 
\ref{FissionBarriers-eta-240U} (b). They display  similar trends as 
functions of the quadrupole moment though, as expected, they become 
larger with increasing $\eta$ values. Concerning the multipole 
moments $Q_{20}(1F)$, $Q_{30}(1F)$, $Q_{20}(2F)$ and $Q_{30}(2F)$ 
shown in panel c), one observes that they lie on top of each other, 
for all the considered $\eta$ values.

In  Fig. \ref{FissionBarriers-eta-240U} (d),  the collective 
inertia  $B_{ATDHFB}$ is depicted. The behavior as a function of the 
quadrupole moment is similar in the three cases but the actual 
values are clearly correlated with the $\eta$ factor. This  is a 
direct consequence of the inverse dependence of the collective mass 
with  the square of the pairing gap \cite{proportional-1,proportional-2}. In particular,  
for $\eta$=1.05 ($\eta$=1.10) the ATDHFB mass is reduced, on the 
average, by 28 $\%$ (46 $\%$). The GCM masses (not shown in the 
plot)  are reduced by 28  and 35 $\%$, respectively. These 
reductions have a significant impact on the predicted fission 
half-lives. For example, for $E_{0}$= 1.0 MeV we have obtained, 
within the ATDHFB scheme, $t_\mathrm{SF}$= 3.215 $\times$ 10$^{42}$, 
3.051 $\times$ 10$^{31}$ and 2.575 $\times$ 10$^{23}$ s for  $\eta$= 
1.00, $\eta$= 1.05 and $\eta$= 1.10, respectively. 

%%%%%%%%%%%%%%%%%%%%%%%%%%%%%%%%%%%%%%%%%%%%%%%%%%%%%%%%%%%%%%%%%%%%%%%%%%%%%%%%%%%%%%%%%%%%%%%%%
%
%   FIGURE OF THE PAPER  
%
%%%%%%%%%%%%%%%%%%%%%%%%%%%%%%%%%%%%%%%%%%%%%%%%%%%%%%%%%%%%%%%%%%%%%%%%%%%%%%%%%%%%%%%%%%%%%%%%%
%%%%%%%%%%%%%%%%%%%%%%%%%%%%%%%%%%%%%%%%%%%%%%%%%%%%%%%%%%%%%%%%%%%%%%%%%%%%%%%%%%%%%%%%%%%%%%%%%

\begin{figure*}
\includegraphics[width=1.0\textwidth]{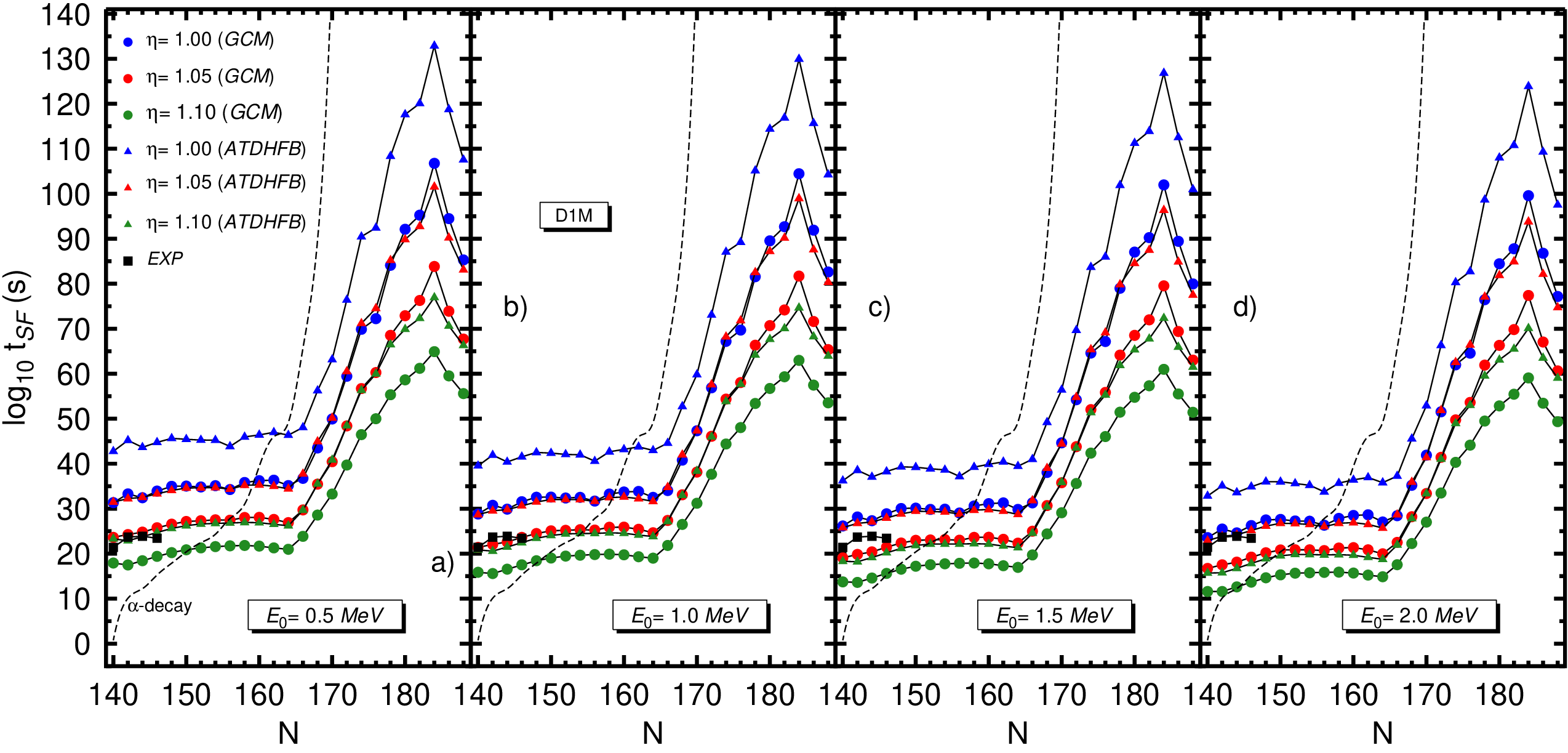} 
\caption{ 
(Color online) The spontaneous fission half-lives $t_\mathrm{SF}$, 
predicted within the GCM and ATDHFB schemes, for the isotopes 
$^{232-280}$U  are depicted as functions of the neutron number. 
Results have been obtained with the normal ($\eta$=1.00) and 
modified ($\eta$=1.05 and 1.10) Gogny-D1M EDFs. Calculations have 
been carried out  with $E_{0}$=0.5 [panel a)], 1.0 [panel b)], 1.5 
[panel c)] and 2.0 MeV [panel d)], respectively. The experimental 
$t_\mathrm{SF}$ values for the nuclei $^{232-238}$U are included in 
the plot. In addition,  $\alpha$-decay half-lives are plotted with  
short dashed lines. For more details, see the main text.
}
\label{tsf-D1M-eta} 
\end{figure*}

In Fig. \ref{tsf-D1M-eta} we have plotted the spontaneous fission 
half-lives $t_\mathrm{SF}$, predicted within the GCM and ATDHFB 
schemes, for the isotopes $^{232-280}$U  as functions of the neutron 
number. Results have been obtained with the normal and modified 
Gogny-D1M EDFs. Calculations have been carried out  with $E_{0}$=0.5 
[panel (a)], 1.0 [panel (b)], 1.5 [panel (c)] and 2.0 MeV [panel 
(d)], respectively. The experimental $t_\mathrm{SF}$ values for the 
nuclei $^{232-238}$U are included in the plot. In addition,  $\alpha$-decay 
half-lives are plotted with  short dashed lines.

 On the one hand, the 
results shown in Fig. \ref{tsf-D1M-eta} illustrate the strong impact 
that pairing correlations have on the fission half-lives in the 
considered Uranium isotopes. Increasing $\eta$, leads to a  
reduction in both $B_{ATDHFB}$ and $B_{GCM}$. As a consequence, for 
a given $E_{0}$, we observe a significant decrease in $t_\mathrm{SF}$
in either the ATDHFB or the GCM schemes. For example, for $E_{0}$= 
0.5 MeV and within the GCM scheme, increasing the pairing strength 
by 5 $\%$ (10 $\%$) leads to a reduction in $t_\mathrm{SF}$ of up to 
9 (16) orders of magnitude in the light isotopes. Such a reduction 
becomes even more pronounced for the  heavier isotopes reaching 23 
(42)  orders of magnitude in the case of $^{276}$U. Note, that our 
results for $^{232-238}$U agree reasonably well with the 
experimental data. However, it is more important that, in spite of 
the large variability in the predicted $t_\mathrm{SF}$ values due to 
pairing correlations, the same global features discussed in the 
previous Sec. \ref{FB-systematcis} (see, Fig. \ref{tsf-D1S-D1N-D1M}) 
still hold: 1) a steady increase in the spontaneous fission 
half-lives is observed for  $N \ge 166$ reaching a maximum at $N=184$; 
2) beyond N=166 the Uranium isotopes can be considered stable with 
respect to spontaneous fission; 3) for increasing  neutron number 
fission turns out to be faster than $\alpha$-decay, with the 
transition point being around $N=144-150$.

Once again, we stress that the results discussed in this and the 
previous Sec. \ref{FB-systematcis}, point to the robustness of the 
overall trend predicted for the spontaneous fission half-lives in 
$^{232-280}$U using the more recent parametrizations of the 
Gogny-EDF. They suggest the use of experimental fission data, 
instead of the more traditional odd-even staggering, to fine tune 
the pairing strengths in those EDFs commonly employed in 
microscopic  studies. They also point \cite{Robledo-Giulliani}, to 
the relevance of beyond mean correlations associated with the 
interplay between pairing fluctuations and particle number symmetry 
restoration \cite{rs} in the description of fission. Given the large 
uncertainties in the predicted $t_\mathrm{SF}$ values, in ours and 
other theoretical approaches, with respect to pairing correlations 
and other building blocks affecting the WKB formula, it becomes 
obvious that a direct comparison with experiment is meaningless. 
Therefore, only the global trends, extracted from calculations 
performed under the same conditions along  series of nuclei and/or 
isotopes, should be used to extract  conclusions.

\section{Conclusions}
\label{conclusions}

In the present work, we have considered the evaluation of fission 
observables within the constrained HFB approximation based on 
Gogny-like EDFs. We have presented a detailed description of the 
methodology employed to obtain the fission paths in the studied 
nuclei. Besides the proton $\hat{Z}$ and neutron $\hat{N}$ number 
operators, we have considered constraints on the axially 
symmetric quadrupole $\hat{Q}_{20}$, octupole $\hat{Q}_{30}$ and 
$\hat{Q}_{10}$ operators. In some instances, we have explored the 
role of the $\gamma$ degree of freedom by means of triaxial 
calculations with simultaneous constraints on both the  $\hat{Q}_{20}$
and $\hat{Q}_{22}$ components of the quadrupole moment. On the 
other hand, HFB solutions corresponding to separated fragments have 
been reached with the help of the necking operator 
$\hat{Q}_{Neck}(z_{0},C_{0})$. The 1F curves obtained in 
this way exhibit a rich topography including the ground state 
minimum, the inner and outer barriers as well as the first and 
second fission isomers. For larger deformations we have found 
2F curves displaying a quasi-linear decrease in energy for 
increasing values of the quadrupole moment. Zero point quantum 
corrections have always been added to each of the mean-field  
solutions {\it{a posteriori}}. In particular, the rotational 
correction has been computed in terms of the Yoccoz moment of 
inertia while two different schemes (i.e., the ATDHFB and GCM ones) 
have been employed in the calculation of both the collective inertia 
and the vibrational corrections. We have thoroughly discussed the 
uncertainties in the predicted spontaneous fission half-lives 
$t_\mathrm{SF}$ arising from different building blocks affecting the 
WKB formula.  

We have carried out Gogny-D1M calculations for a selected set of 
actinides and superheavy elements. The comparison between the 
theoretical and experimental inner and  second barrier heights as 
well as  the excitation energies of fission isomers shows that the 
global trend observed in the experiment is reasonably well 
reproduced. The same is true in the case of the spontaneous fission 
half-lives, regardless of whether the ATDHFB or GCM masses are used. 
In particular, our results demonstrate that the Gogny-D1M HFB 
framework captures the severe experimental  $t_\mathrm{SF}$ 
reduction between $^{232}$U and $^{286}$Fl as well as the trend 
along different isotopic chains. Another relevant source of 
information is the mass and charge of the resulting fission 
fragments, which are determined by the nuclear shape in the 
neighborhood of the scission point. In our calculations the proton 
and neutron numbers of the fragments are determined by energetic 
considerations and therefore they are mostly dominated by the $Z=50$ 
and $N=82$ magic numbers. Those values, however, underestimate by 
several mass units the experimental values pointing to the need 
of a better dynamical theory to describe post-fission phenomena. The 
results obtained validate the D1M Gogny-EDF, originally 
tailored to better reproduce nuclear masses, for the study of 
fission properties in heavy and superheavy nuclei.

We have performed a systematic study of the fission properties in 
Uranium nuclei, including very neutron-rich isotopes up to $^{280}$U. 
In order to verify the robustness of our predictions, when 
extrapolated to very exotic N/Z ratios,  calculations have been 
carried out with the three most recent incarnations of the 
Gogny-EDF, i.e., the parametrizations D1S, D1N and D1M. The well 
known under-binding  of the heavier isotopes characteristics of  the Gogny-D1S EDF is 
clearly visible in our calculations. Nevertheless, the fission paths 
still exhibit rather similar shapes regardless of the functional 
employed. An increase in the height of the inner fission barriers 
and the widening of the 1F curves appear as  common 
features as we approach the two-neutron dripline. Second fission 
isomers are predicted for several  Uranium isotopes. From the 
systematics of the spontaneous fission  half-lives we conclude 
that,  even when subject to large uncertainties, the Gogny HFB 
framework produces a trend which is quite robust. In particular, we 
have found that: 
\begin{itemize} 
\item a steady increase in the spontaneous fission half-lives is 
observed for  $N \ge 166$ with a peak at the neutron magic number $N=184$ 
\item beyond $N=166$ the Uranium isotopes can be considered stable with respect to 
spontaneous fission 
\item as a decay mode fission becomes faster 
than $\alpha$-emission for increasing  neutron number. 
\end{itemize} 
In addition, the analysis of the masses and charges of the fission 
fragments reveals, the key role played by the $Z=50$ and $N=82$ 
shell closures in the splitting of the considered Uranium isotopes. 
Interesting enough, oblate deformed fragments are predicted in our 
calculations that deserve further attention as it is usually assumed 
that fission fragments exhibit prolate deformations.

In the present study special attention has been paid to the impact 
of pairing correlations on the  fission properties in $^{232-280}$U. 
To this end, we have also considered a modified Gogny-D1M EDF in 
which the pairing strengths  are increased by 5 and 10 $\%$, 
respectively. On the one hand, our calculations further corroborate 
the robustness of the predicted spontaneous fission half-lives 
systematics. On the other hand, they also  illustrate that 
modifications of such a few per cent in the pairing strength can 
have a dramatic impact on the collective masses therefore altering 
the absolute values of the fission half-lives by several orders of 
magnitude. Within this context, we advocate the  use of experimental 
fission data, instead of the more traditional odd-even staggering, 
to fine tune the pairing strengths in those EDFs commonly employed 
in microscopic  studies. Our results also point, to the relevance of 
beyond mean correlations associated with the interplay between 
pairing fluctuations and particle number symmetry restoration in the 
description of nuclear fission. 

Last but not least, let us also comment on a more methodological 
issue. Given the present state of affairs in the microscopic 
computation of spontaneous fission half-lives, even in the case of 
state-of-the-art approximations, it is highly desirable to explore 
new avenues in which the minimization of the action S [see, Eq. (
\ref{Action})] acquires a central role. The first steps, within the 
Skyrme-EDF framework, have been undertaken very recently 
\cite{ACTION-DOBA}. The action is proportional to the square root of the 
collective inertia and therefore any degree of freedom having an 
impact on it, will play an essential role. In this 
respect, pairing correlations should be incorporated as an important 
degree of freedom, in addition to the more traditional quadrupole 
and octupole moments. Work along these lines is in progress and will 
be reported elsewhere.

\begin{acknowledgments}
Work supported in part by MICINN grants Nos. FPA2012-34694, 
FIS2012-34479 and by the Consolider-Ingenio 2010 program MULTIDARK 
CSD2009-00064. This work was completed while one of the authors 
(LMR) participated at the INT13-3 program. The warm hospitality of 
the Institute for Nuclear Theory and the University of Washington is 
greatly acknowledged.
\end{acknowledgments}

\end{document}